# An automated pipeline for computation and analysis of functional ventilation and perfusion lung MRI with matrix pencil decomposition: TrueLung

**Authors:** Orso Pusterla[1,2,3*], Corin Willers[3], Robin Sandkühler[2], Simon Andermatt[2], Sylvia Nyilas[4], Philippe C. Cattin[2], Philipp Latzin[3], Oliver Bieri[1,2], and Grzegorz Bauman[1,2]

**Author Affiliations**

[1]*Department of Radiology, Division of Radiological Physics, University Hospital Basel, University of Basel, Basel, Switzerland*

[2]*Department of Biomedical Engineering, University of Basel, Basel, Switzerland*

[3]*Division of Pediatric Respiratory Medicine and Allergology, Department of Pediatrics, Inselspital, Bern University Hospital, University of Bern, Switzerland*

[4]*Department of Diagnostic, Interventional and Pediatric Radiology, Inselspital, Bern University Hospital, University of Bern, Switzerland*

**\* Corresponding author:**

Orso Pusterla, PhD

Division of Radiological Physics, University Hospital Basel, Petersgraben 4, 4031 Basel, Switzerland

E-mail: orso.pustrerla@unibas.ch ; Phone: +41.76.693.02.20;

ORCID-ID: https://orcid.org/0000-0003-1879-055X

**Paper details:** Abstract 300 words. Paper Body 4800 words, 8 Figures, 0 Tables, 1 Appendix.

Supplementary Material for review and online publication: 1 Supplementary PDF document.

**ABSTRACT**

**Purpose:** To introduce and evaluate TrueLung, an automated pipeline for computation and analysis of free-breathing and contrast-agent free pulmonary functional magnetic resonance imaging.

**Materials and Methods:** Two-dimensional time-resolved ultra-fast balanced steady-state free precession acquisitions are transferred to TrueLung, which includes image quality checks, image registration, and computation of perfusion and ventilation maps with matrix pencil decomposition. Neural network whole-lung and lobar segmentations allow quantification of impaired relative perfusion ($R_Q$) and fractional ventilation ($R_{FV}$). TrueLung delivers functional maps and quantitative outcomes, reported for clinicians in concise documents.

We evaluated the pipeline using 1.5T images from 75 cystic fibrosis children. Whole-lung and lobar segmentations were manually refined when necessary, and the impact on $R_Q$ and $R_{FV}$ was quantified.

**Results:** Functional imaging was performed at 7.9±1.8 (mean±SD) coronal slice positions per patient, totaling 6min 20s scan time per patient. The whole pipeline required 20min calculation time per subject. TrueLung delivered the functional maps of all the subjects for radiological assessment. Quality controlling maps and segmentations lasted 1min 12s per patient. The automated segmentations and quantification of whole-lung defects were satisfying in 88% of patients (97% of slices) and the lobar quantification in 73% (93% of slices). The segmentations refinements required 16s per patient for the whole-lung, and 2min 10s for the lobe masks.

The relative differences in $R_{FV}$ and $R_Q$ between fully-automated and manually refined data were 0.7% (1.2%) and 2.0% (2.9%) for whole-lung quantification (median, [third quartile]), and excluding two outliers, 1.7% (3.9%) and 1.2% (3.8%) for the lobes, indicating the refinements could be potentially omitted in several patients.

**Conclusions:** TrueLung quickly delivers functional maps and quantitative outcomes in an objective and standardized way, suitable for radiological and pneumological assessment with minimal manual input. TrueLung can be used for clinical research in cystic fibrosis and might be applied across various lung diseases.

**1. INTRODUCTION**

To date, pulmonary diseases remain one of the most significant public health challenges causing morbidity worldwide and severely limiting the quality of life in both young and adult populations [1]. Nevertheless, disease characterization and early interventions might slow down the rate of lung function decline [2]. Hence, methods to detect and quantify subclinical lung function impairments even at early stages as well as to monitor the therapeutic responses are crucial to improve patients' care. Accordingly, there is a need for sensitive biomarkers and clinically relevant intermediate outcomes to improve understanding of pulmonary disease evolution and develop optimal targeted therapies.

Pulmonary function tests (PFTs) – such as spirometry, plethysmography, multiple-breath washout measurements, and diffusing capacity of the lungs for carbon monoxide – provide global parameters of disease severity and have shown compelling clinical relevance [3]. However, the diagnostic value of PFTs is limited as they do not provide insight into precise morphological changes and do not quantify lung function regionally or show undergoing pathophysiological mechanisms [4]. In addition, if deficits affect only exiguous parts of the lung, healthy portions could compensate, resulting in normal or only slightly abnormal PFTs outcomes. Therefore, PFTs might not be sensitive enough to detect the disease in its early stages and the progression [3–5], overall limiting the opportunities for prompt and personalized interventions.

To overcome these limitations, several medical imaging modalities have been developed for regional assessment of lung morphology and function, allowing identification of disease phenotypes and improvement of therapies. Thanks to its diagnostic capability, imaging nowadays plays a major role in assessing pulmonary diseases and is considered a mainstay of contemporary respiratory medicine.

The current clinical gold standards for morphological and functional assessment of the lung are computed tomography (CT) and nuclear medicine imaging modalities [6]. However, these imaging modalities carry the risk of potentially harmful ionizing radiation. This is of particular concern and contraindicated for serial imaging, especially in pediatric subjects requiring frequent monitoring of the lung development or disease progression, or response to therapeutic procedures [7,8]. Moreover, exposure to ionizing radiation should be avoided, if possible, in pregnant women or for research investigations.

A viable solution is offered by magnetic resonance imaging (MRI), which does not involve ionizing radiation [9] and is thus well suited for frequent examinations of the lung structure and function [2,10–14]. Nowadays, several dedicated MR pulse sequences offer relevant clinical imaging for morphological examinations [11,12,15–17]. Within the last decades, particular focus was also put on pulmonary biomarkers related to ventilation, gas exchanges and perfusion [5,18–20], since functional abnormalities might be observable before any structural change [21]. For example, reduced regional lung ventilation and decreased perfusion due to subsequent hypoxic pulmonary vasoconstriction could be caused by mucus obstructions of small airways, which cannot be directly visualized by either morphological MRI or high-resolution CT. Furthermore, early pathology in peripheral lung compartments and reduced functions might represent a precursor of irreversible lung damage [22,23].

MRI with inhaled gas agents (e.g., 3He, 129Xe, and 19F) has shown potential for ventilation and gas-exchange imaging, while dynamic contrast-enhanced (DCE) imaging using intravenous gadolinium-based contrast agents is the most established MR technique for pulmonary perfusion assessment [10,11]. However, these techniques have some drawbacks. MRI with gas agents necessitates very specialized equipment and personnel. DCE requires intravenous injection of contrast agent, increasing complications for imaging, patient discomfort, and posing health risks with associated rare allergic reactions and rare nephrogenic systemic fibrosis [24]; moreover, there are concerns regarding contrast agent deposition in the body. It has to be stressed that concerns about contrast-enhanced and ionizing radiation imaging should not lead to a denial of well indicated and pivotal radiological examinations with DCE-MRI, CT, and nuclear medicine modalities. However, these modalities should be prescribed only when crucially necessary for a clinical diagnosis and there are no viable alternatives [7,8]. On the other hand, they are not advisable especially for children and babies requiring periodic and frequent measurements for monitoring lung development.

Hence, simple and non-invasive MRI techniques for assessing regional pulmonary functions are highly desirable. In the last decade, proton-based pulmonary functional MRI with Fourier decomposition (FD) postprocessing [25] and derivate approaches such as matrix pencil (MP) [26], PREFUL [27], SENCEFUL [28], and dynamic mode decomposition (DMD) [29], have been proposed for simultaneous assessment of regional lung ventilation and perfusion with promising results in several lung pathologies [5,19,27,30–33]. These techniques rely on free tidal-breathing time-resolved MR acquisitions of the lung and the naturally occurring respiratory and cardiac signal modulations in the pulmonary tissue, without any need for neither administration of intravenous contrast agents nor inhaled tracers. The lung expands and contracts at every breath, and at every heartbeat, the blood is pumped through the pulmonary arteries to reach the lung parenchyma. This creates the signal modulations used to compute maps of fractional ventilation, perfusion, and blood arrival time. Hence, pulmonary functional MRI can provide visual as well as quantitative information about functional deficits of the lung.

Matrix pencil decomposition is an improved variant of FD-MRI, which increases the accuracy and robustness of the results mitigating the problem of irregular breathing patterns and cardiac frequencies, as well as time-series truncation. MP-MRI employs dedicated and specifically designed 2D time-resolved imaging sequences allowing for significant improvement of lung imaging quality, which is fundamental for reliable functional imaging [34,35].

Previous studies have demonstrated the potential of MP-MRI to detect functional impairments in various lung diseases, from early childhood on and even in toddlers [19,30,36–38]. MP-MRI has also been proven initial validation against gold-standard imaging modalities, including the clinical gold-standard single-photon emission computed tomography (SPECT), DCE, and inhaled hyperpolarized gas MRI [37,39,40]. The reproducibility of MP-MRI was investigated and strong correlations with PFTs outcomes were found [30]. Figure 1 shows representative fractional ventilation and perfusion maps as well as regions with impaired lung functions computed with MP-MRI in two children with cystic fibrosis (CF), one clinically healthy and one with severe lung impairments.

MP-MRI data can be acquired on a standard clinical whole-body scanner and its feasibility was demonstrated at 0.55T, 1.5T, and 3T [34,35,41]. Overall, simple and free-breathing MP-MRI is highly promising, and an easy, advantageous, and broad integration into many clinical centers is desired and possible. Nevertheless, the computation of MP functional maps and quantitative outcomes requires several specific postprocessing steps, which include image registration [42], computation of ventilation and perfusion maps with the MP algorithm [26], lung segmentation [43–45], and calculation of functional defects [19]. In addition, the MP-MRI functional maps, the outcomes, and the reports must be quickly available to the treating physicians for clinical decision-making and case discussion with the patients. Consequently, to facilitate broad clinical and research applications, the pipeline of postprocessing steps must be automated, fast, robust, and reproducible. Ideally, objective and standardized quality control is also included. Further, for a widespread of the technique, due to the confidentiality of the data, the pipeline must be secure, and implemented to run locally on a standard workstation.

Previously, a framework for Fourier-decomposition ventilation imaging was implemented by Guo et al. [46], but their methodology was limited to ventilation imaging only, i.e. no perfusion-weighted maps were presented. Recently, Crisosto et al. also, in their preliminary research, proposed a framework for PREFUL MRI [47]. Still, their pipeline requires a super-computer or prohibitively long computation times on a standard workstation. Moreover, these frameworks did not use the MP algorithm, specific image registration, which considers the lung-ribcage sliding motion [42], advancements in whole-lung and lung-lobe segmentations with artificial neural networks [45], and no evaluations on a large cohort were performed.

In this study, we introduce and provide a detailed description of the automated pipeline TrueLung, which includes all processing steps from image acquisition to the delivery of clinical reports, enabling both ventilation and perfusion outcomes even at the lobar level. Our primary objective is to evaluate the TrueLung pipeline on a cohort of 75 children with CF, aiming to assess its accuracy and effectiveness in analyzing lung function.

## 2. MATERIALS AND METHODS

### 2.1. The TrueLung pipeline

A schematic of the automated TrueLung pipeline and the end-to-end workflow starting from image acquisition until the delivery of clinical reports is presented in Figure 2. The pipeline is described here, and specific details are described later in Appendix A.

Functional imaging for MP-MRI consists of two-dimensional time-resolved (2D+t) coronal acquisitions (base images) during less than one minute of simple free tidal-breathing per slice. Imaging is usually performed at 6 to 10 equidistant coronal slice positions to cover the majority of the lung volume, with the specific number of slices depending on the size of the subject; this typically results in an overall scan time of about 5 to 9 minutes. Typically, an ultra-fast balanced steady-state free precession (ufSSFP) pulse sequence is employed at 1.5T and 0.55T, while at 3T a dedicated transient spoiled gradient echo (tSPGR) sequence is used, and discussed later [34,35,41]; in this section, for simplicity in wording, we refer to the ufSSFP sequence only.

The MR acquisitions are stored in the PACS (picture archiving and communication system) local server and can be accessed by radiologists. The ufSSFP datasets are transferred from the PACS or the MR scanner to a local workstation equipped with a CUDA (Compute Unified Device Architecture) (NVIDIA, Santa Clara, CA) compatible GPU on which TrueLung is installed, for example within a Docker container.

The TrueLung pipeline reads the coronal ufSSFP data and sorts them from anterior to posterior slice position, then the data are processed slice by slice. The first ten images of every time series are excluded from the processing since acquired in the transient magnetization state [34]. As the first quality-check, an analysis of the whole-lung signal modulations related to ventilation is performed using the matrix pencil algorithm to identify possible unwanted patient movement, coughing, extremely shallow or no breathing. In these cases, the whole or part of the data not following ventilation-related sinusoidal signal modulations are discarded, the data are flagged, and a warning message is printed into a logfile. The data passing the quality-check are further analyzed with the matrix pencil algorithm to identify images at three respiratory phases automatically. The first image is chosen in a middle respiratory phase and serves as baseline image to register the time-series. The second and third images are in expiration and inspiration, and are segmented via an artificial neural network: the lung areas as calculated on segmentation masks are used to compute the fractional ventilation per slice, employed in a second moment to verify the per slice fractional ventilation computed with the MP algorithm.

The 2D ufSSFP image time-series are registered to the fixed images chosen in the mid-respiratory state (baseline images). The image registration is performed using a self-developed algorithm that preserves ventilation and perfusion signal modulations, aligns automatically moving lung structures such as vessels and airways, but preserves non-moving structures such as the thoracic cage [42] by considering the discontinuities in the motion field automatically.

Subsequently, the whole lung and the lung lobes are segmented with an artificial neural network, required for automated defect quantification [45]. Finally, the registered time-series are processed voxel-wise with the MP algorithm [26] to calculate maps of fractional ventilation, perfusion, and blood arrival time (not shown in Figure 2). Here the MP algorithm uses the lung segmentations for precise spectral analysis of the frequencies related to ventilation and perfusion. The previous lung segmentation of inspiration/expiration images are used to verify the fractional ventilation values computed with MP. Lung vessels are segmented on perfusion maps, and perfusion values quantified. All the functional maps can be transmitted back to the PACS or an independent server for radiological examinations.

Previously obtained lung segmentations from inspiration and expiration images are employed to validate fractional ventilation values derived using MP. This involves comparing the fractional ventilation values computed through segmentations with those obtained via MP

In a final step, an algorithm (e.g. histogram distribution analysis) allows quantifying whole-lung and lobar perfusion and ventilation defects, and creating masks of pulmonary functional defects. For case discussions and decision-making by pneumologists, the main quantitative outcomes for the whole lung and for every lung lobe are summarized in a comprehensive PDF report together with lung functional maps and lung defect maps. The functional defect values of every slice are saved in an accessible tabulated text file. All the functional maps, the defect maps, the text file, and reports are saved. In this final step, which includes automated functions quantification, whole-lung and lung-lobe segmentation masks are required and must be quality controlled by a human observer; if necessary, the segmentation masks are manually refined, and both the MP computations and the function quantification steps are repeated. Finally, the reports and the functional maps are securely sent to the referring physician for interpretation and to PACS.

An in-depth overview of the TrueLung pipeline is presented in Appendix A, covering the following aspects: imaging for TrueLung, data transfer, image registration, recurrent neural network segmentation, the MP algorithm, quantitative evaluation and outcomes, reporting, and quality control.

**2.2 Pipeline evaluation: study design, data acquisition and analysis**

In this cross-sectional study, data from 75 children with CF were included. The CF study population represents a broad spectrum of the pediatric population from 4 to 19 years (12.4±4.5 years old, mean ±SD). The lung clearance index (LCI) ranged between 7.1 to 18.2 turnovers, and the forced expiratory volume in 1 second (FEV-1) between 1.2 to -5.8 z-scores.

The children underwent a comprehensive contrast-agent free MRI study protocol for morphological and functional pulmonary assessment. For functional assessment of the whole lung with ufSSFP, coronal images were acquired at several equidistant anterior-posterior slice locations and automatically processed with the TrueLung pipeline. For this methodological work, data from single-center cross-sectional observational studies at the University Children's Hospital of Bern, Switzerland were used [19,30,44]. This study and all the measurements were conducted in agreement with the local ethic regulations. Written informed consent was obtained by parents and by participants if older than 14 years.

Imaging was performed on a 1.5T whole-body MRI Scanner (MAGNETOM Aera, Siemens Healthineers, Erlangen, Germany). Children were not sedated during the scans. Two-dimensional time-resolved ufSSFP coronal sets were acquired during 48 seconds of free-breathing [34]. Scan parameters for ufSSFP were as follows: field-of-view 425×425 mm$^2$, matrix size 128×128 interpolated to 256×256 (native in-plane resolution $3.32 \times 3.32$ mm$^2$, interpolated to $1.66 \times 1.66$ mm$^2$), slice thickness 12 mm, echo time (TE) 0.67 ms, repetition time (TR) 1.46 ms, bandwidth 2056 Hz/pixel, flip angle 65°, GRAPPA factor 2, 160 coronal images per slice. The nominal acquisition time for one image was 120 ms, followed by a waiting time of 180 ms, resulting in a total acquisition time of 300 ms per slice and an acquisition rate of 3.3 images per second.

A threshold equal to 75% of the median value from each voxel distribution was used to quantify regions with impaired lung ventilation ($R_{FV}$) and perfusion ($R_Q$) in every slice, as described before [19,44]. Quantification was performed for the whole-lung and lung lobes. The lung lobes were segmented with the RNN as explained before [45], and the union of all the lobes was used as whole-lung segmentation mask. None of the examinations evaluated in this study were included in the RNN training data; the TrueLung evaluation in this study represents, thus, a new and independent testing set for the network.

As main study outcomes, we measured the scan duration and the number of acquired ufSSFP slices per patient. We measured the time required for image registration, segmentation, analysis with MP, quantification of functional defects, the total time required for the whole TrueLung pipeline, quality controlling and manual interventions. Moreover, we evaluated the percentage of datasets that TrueLung did not process perfectly due to segmentation flaws. The impact of segmentation imperfections on $R_{FV}$ and $R_Q$ was aalyzed with Bland-Altmann analysis between datasets fully automatically analyzed and the data quality controlled and whose segmentations were manually corrected. The 95% limits of agreement (LOA) were calculated as the mean difference ±1.96 SD. Similarly, the coefficient of reproducibility (RPC) as 1.96 times the standard deviation of the differences. In addition, the median and third quartile (q3/4) of the absolute relative error in $R_{FV}$ and $R_Q$ between the datasets fully automatically analyzed and the one which required manual segmentation refinement was calculated. We run the pipeline twice to evaluate its reproducibility.

The TrueLung pipeline and the analysis were performed on a Linux workstation equipped with an eight cores CPU (Intel Xeon E5-1680, 3 GHz, Santa Clara, CA), 64GB RAM, a solid-state drive (SSD), and a CUDA compatible GPU (NVIDIA GeForce RTX 2070, 8 GB RAM, Santa Clara, CA).

## 3. RESULTS

MR functional imaging was successfully completed by all the 75 children with CF included in this study without any dropouts. Functional imaging was acquired at 7.9±1.8 (mean±SD) slice locations per patient, resulting in a mean scan time of approximately 6 minutes and 20 seconds per patient. The time required for image registration was 17min 6s per patient, for segmentation 2min 10s, for the MP analysis 30s, and for defect quantification and generation of reports less than a second. The whole TrueLung pipeline required 20 minutes to process all the data of one patient, end-to-end.

Representative baseline image, functional maps, segmentation masks, and functional defects masks obtained with TrueLung in a patient with CF showing severe lung disease are presented in Figure 3. An exemplary shortened PDF report of the main outcomes is given in Figure 4 for the same patient with CF, whereas the full PDF report is given in the Supplementary Material S1 available online. The detailed outcome parameters divided per slices, useful for clinical research, are printed in a tabulated text file, shown in Supplementary Material S2, available online. The CF patient presented in Figure 3 and Figure 4 has a conspicuous ventilation and perfusion defects in the right upper lobe and left lower lobe, as demonstrated with TrueLung, which allows for the regional assessments and quantification.

The whole-lung and lobar segmentation masks overlaid on the baseline images (see Figure 3) for the 75 patients were quality controlled in circa 90 minutes (1min 12s per patient). Representative well-executed whole-lung and lobar segmentations, as well as masks which were manually refined, are exemplarily shown in Figure 5.

Figure 6 presents a flowchart of TrueLung success rate for the 75 patients processed. All the 75 patients (589 slices) could be processed by TrueLung, and the functional maps were delivered to radiologists for findings. No unwanted patient movement was detected. Regarding the whole-lung quantitative evaluation of defects ($R_{FV}$ and $R_Q$), in 66 patients, the TrueLung pipeline performed perfectly (574 slices). In 9 patients the RNN whole-lung segmentation had minor flaws (in a total of 15 slices), and the masks had to be refined. The manual correction of the 15 whole-lung masks required about 20 minutes.

The lobar segmentations were well executed for 55 patients (549 slices). In 20 patients a total of forty slices required minor manual refinements due to the mixing of lobe boundaries (see Figure 5). The manual correction of the lobar segmentation masks required about 4 minutes per slice.

After manual correction of the whole-lung and lobar masks, the datasets of these patients were reprocessed by the MP module in TrueLung and reports of quantitative pulmonary functions were generated (cf. Figure 5) and sent to clinicians.

Overall, in the collective of patients, the TrueLung pipeline delivered all functional maps (100% success rate). For whole-lung defect evaluation, the pipeline was able to process without the need for manual intervention 97% of slices (574/589) correctly, or from another perspective, 88% (66/75) of the patient examinations. The lobar defect evaluation was well performed in 93% of slices (549/589), or from a different viewpoint, in 73% (55/75) of patient examinations. The manual refinements of whole-lung and lobar segmentation masks took 16s and 2min 10s per patient, respectively.

Bland-Altman plots showing the variability of V, and Q relative defect percentage ($R_{FV}$ and $R_Q$) in the whole lung for datasets fully automatically analyzed, and the quality-controlled datasets whose segmentations were manually corrected if required, are presented in Figure 7. As corroborated by the tights LOAs and RPCs, the refinements of segmentation masks change only marginally the whole-lung functional outcomes, except for two obvious outliers, whose segmentation masks are presented in Figure 7. Excluding the two outliers, the median and q3/4 absolute relative errors between automatic and manually corrected data (7 patients) are 0.7 (1.2, range from 0.2 to 1.9) [%] for $R_{FV}$, and 2.0 (2.9, range from 0.2 to 4.9) [%] for $R_Q$.

Bland-Altman plots of lobar $R_{FV}$ and $R_Q$ outcomes for datasets fully automatically analyzed and the quality-controlled datasets whose segmentations were manually corrected if required, are presented in Figure 8. The two outliers previously discussed (Figure 7) were removed from the Bland-Altman analysis and hereafter. Also for the lobar analysis, refinements of lobar masks affect only marginally the functional outcomes. Excluding the two outliers, the median and q3/4 absolute relative error between automatic and corrected data (20 patients) for the upper left lobe, lower left, upper right, middle right, and lower right, are 1.9 (6.1), 1.4 (1.8), 1.2 (2.7), 3.1 (4.6), 0.7 (4.5) [%] for $R_{FV}$, and 1.1 (5.4), 1.4 (4.4), 0.6 (1.8), 1.9 (2.7), 0.9 (4.8) [%] for $R_Q$. The all-lobe median and q3/4 absolute relative error were 1.7 (3.9) [%] for $R_{FV}$ and 1.2 (3.8) [%] for $R_Q$.

The reproducibility of TrueLung was evaluated by running the whole pipeline twice on all 75 subjects, and it was perfect, i.e., there were no variations.

## 4. DISCUSSION

For monitoring pulmonary diseases sensitive biomarkers and clinically relevant outcomes are required. These measures might be provided by structural and functional MRI. Generally, there is an increased interest in imaging biomarkers related to pulmonary functions because they might be more sensitive to early stages of diseases as compared to structural changes. To this end, fast, free-breathing, contrast-agent-free 1H functional imaging is very attractive since it provides unique functional information and can be broadly applied in clinics without the need for specific hardware. Nevertheless, dedicated imaging and computational demanding post-processing steps are needed. In this study, we proposed and evaluated the TrueLung pipeline for the automated computation of functional maps and quantification. The pipeline works fast and efficiently on a dedicated workstation on a Docker container and processes the data of one subject in about 20 minutes.

The feasibility of TrueLung was demonstrated in a collective of young CF patients scanned at 1.5T, and the pipeline had a 100% success rate in delivering functional maps to radiologists. The whole-lung pulmonary biomarkers could be quantified in 88% of patients automatically, whereas the remaining 12% of patient examinations were successfully processed after quick manual refinement of the required whole-lung segmentation masks. The automated lobar evaluation of biomarkers, which increases the complexity of the quantification but also prospects for more sensitive outcomes, was perfect in 73% of patients, whereas for 27% of patients, the evaluation was recalculated after lobar masks refinements. Overall, only 3% of whole-lung masks and 7% of lobar masks were refined, and partly only minimally. Consequently, the manual intervention changed the patient outcomes only very marginally, except for two outlier cases that were easily recognizable (see Figure 7). Hence, the masks refinements could have been neglected in the majority of cases, potentially increasing the efficiency of TrueLung to in analyzing patient examinations to about 97% (73/75). Further evidence is given by the exiguous RPCs of $R_{FV}$ and $R_Q$ between automated and manually corrected data (absolute difference <0.8% for whole-lung quantification), which are smaller than the RPCs of a reproducibility study (<4.1%) in which MR measurements were repeated 24h apart [30]. Moreover, the RPCs in $R_{FV}$ and $R_Q$ of our work are similar or lower to the RPCs found in an intra-observer repeatability (<0.9%) and inter-observer reproducibility study (<2.3%) in which whole-lung segmentations were outlined twice by the same observer and by two different clinicians [44]. Excluding obvious outliers, overall the impact of marginal segmentation flaws is thus negligible for both cross-sectional single-point and longitudinal assessments. The reproducibility of TrueLung was perfect, as expected, since all computational steps are deterministic.

Previous works have shown the potential for automatizing the postprocessing for 1H functional imaging [46,47]. In their work, Guo et al. were able to process the dataset of one patient in about 90min (considering nine coronal slices), but their method was restricted to ventilation only. On the other hand, Crisosto et al., in their preliminary research were able to provide both ventilation and perfusion functional information, but their method requires about 8 hours to evaluate the datasets of one patient consisting of 8 slices. The proposed TrueLung pipeline delivers both perfusion and ventilation information in approximately 20 minutes, facilitating decision-making, timed clinical interventions, and communication with patients. Further originality of our work is the evaluation of the whole pipeline performance in a large collective of patients. The postprocessing of TrueLung was accelerated thanks to the image registration algorithm able to run on a GPU and multi-core MP computations. Another novelty of our work is the image registration algorithm that was specifically designed for lung imaging [42] and is able to mitigate the problem caused by sliding organs, such as in the thorax. Moreover, thanks to RNN segmentations, our pipeline is able to provide not only quantitative evaluation of the whole-lung and for every slice, but uniquely, also divided per lung lobes. Being composed by several automated modules and thanks to its speed, TrueLung offers an interesting platform for clinical research since each component can be modified and the whole pipeline re-run effortless, if necessary. For example, novel outcome parameters such as V/Q overlap, the recently proposed defect distribution index [48] or more advanced defect quantifications such as neural networks-driven one [49–51], can be easily integrated, and entire study populations re-evaluated.

Minimal input is still required for quality controlling the processed datasets used for quantitative outcomes (1min 12s per patient) and is unavoidable since validation and decisions are made by humans. The automated quantitative evaluation did not reach a 100% success rate due to improper RNN segmentations. These errors were easily recognized in the maps, and the masks were corrected manually in little time. We remark that, similarly to many artificial neural networks for segmentation tasks, inaccuracies still happen, but the impact of few flawed segmentation on quantitative parameters is almost negligible, as corroborated by our results. Nevertheless, there is room for further improvements. We could address this by including into the neural network recent advancements such as attention layers, or further training the network, for example, with a federated learning approach to benefit from multicenter data without concerns for data anonymity and handling [52].

In this study, TrueLung was evaluated using coronal data acquired at 1.5T with the custom-developed ufSSFP sequence on a single MR scanner. While TrueLung can process MR data acquired with commercially available SSFP and SPGR sequences at various field strengths and on scanners from different vendors, automated segmentation may be less accurate due to differing signal intensity profiles compared to 1.5T. Nevertheless, TrueLung can still compute pulmonary perfusion and ventilation maps for radiological assessment even without the automated segmentation module (i.e., without automated quantification and reporting) and also in sagittal orientation, enabling broader clinical utility. Considering the demonstrated benefits of ufSSFP at 1.5T and 0.5T, and tSPGR at 3T for pulmonary functional imaging [34,41], the implementation of vendor-agnostic versions of these custom pulse sequences, possibly through platforms like pulseq [53], would further broader clinical research. Additionally, specific adjustments may be required in TrueLung to ensure compatibility with DICOM data from scanners of different vendors, particularly for data export and import to PACS and processing.

Our work focused on investigating TrueLung's feasibility in a cohort of children aged 4-19 years old. Based on findings from previous studies [54], we hypothesize that TrueLung might yield promising results in adult subjects as well. On the other hand, imaging for toddlers and neonates should be optimized to account for their rapid cardiac frequency; for example, by speeding up the acquisition rate, reducing the waiting time between acquisitions.

Although current results of proton-based functional lung MRI in general [26–29] and MP-MRI specifically are very promising, broader clinical application and research are still required to demonstrate the full potential of the techniques. Nevertheless, the outcomes need to be quickly and easily available to investigate MRI biomarkers against other methods and incorporate them into clinical decision-making. We addressed the drawback of the time-consuming processing with TrueLung, which can be simply used with minimal and simple user input. TrueLung might allow thus wide clinical rollout in several lung diseases affecting lung function. In addition, there are further aspects to be investigated regarding functional MRI and there is potential for research. The quantification of functional defects can be performed using several algorithms [48–51], but the best clinical outcomes might be yet to be identified and might depend on the disease investigated. Moreover, the lobar quantification offered by TrueLung might provide novel dimensions for more accurate analyses, but it still necessitates clinical studies to prove its potential.

## 5. CONCLUSIONS

We developed and evaluated TrueLung for automated pulmonary functional MRI data processing and analysis, demonstrating excellent results in young patients with CF. TrueLung might accelerate the clinical transition to MR, and we foresee potential for broad application in several lung diseases.

**Acknowledgements**

Orso Pusterla acknowledges the support of the Swiss Cystic Fibrosis Society (CFCH). The CFCH had no role in study design, data collection and analysis, decision to publish, or preparation of the manuscript.

**Declaration of interests**

The authors declare that they have no known competing financial interests or personal relationships that could have appeared to influence the work reported in this paper.

**Data Availability**

Upon request, the TrueLung pipeline will be available to interested researchers or clinicians for collaborations.

**CRediT authorship contribution statement**

Orso Pusterla: Conceptualization, Methodology, Software, Validation, Formal analysis, Formal analysis, Investigation, Data Curation, Writing, Writing-Review & Editing, Visualization, Supervision, Funding acquisition. Corin Willers: Conceptualization, Methodology, Validation, Investigation, Data Curation, Writing-Review & Editing. Robin Sandkühler: Conceptualization, Methodology, Software, Writing-Review & Editing. Simon Andermatt: Methodology, Software, Writing-Review & Editing. Sylvia Nyilas: Methodology, Investigation, Resources, Data Curation, Writing-Review & Editing, Funding acquisition. Philippe P. Cattin: Conceptualization, Methodology, Resources, Writing-Review & Editing, Supervision, Project administration, Funding acquisition. Philipp Latzin: Conceptualization, Methodology, Resources, Writing-Review & Editing, Supervision, Project administration, Funding acquisition. Oliver Bieri: Conceptualization, Methodology, Software, Resources, Writing-Review & Editing, Supervision, Project administration, Funding acquisition. Grzegorz Bauman: Conceptualization, Methodology, Software, Investigation, Resources, Data Curation, Writing-Review & Editing, Supervision, Project administration, Funding acquisition.

## APPENDIX A: In-depth overview of the TrueLung pipeline

### A.1. Imaging for TrueLung

Proton-based pulmonary imaging is notoriously challenging due to the physical properties of the lung (i.e. low proton density, very short T2 and T2* and long T1 relaxation times) and the continuous respiratory and cardiac motions. Substantial progress has been made thanks to optimized acquisition schemes for improved visualization of pulmonary tissue and functions recently.

For MP-MRI two-dimensional time-resolved image series are needed and are acquired with dedicated and optimized MR pulse sequences. At 1.5T and 0.55T imaging is performed with an ufSSFP pulse sequence [34,41], while at 3T with a tSPGR sequence [35]. Due to the very short T2 and T2* relaxation times of lung tissue [34], the key factors for both ufSSFP and tSPGR for pulmonary imaging are very short echo time (TE) and repetition time (TR), which are achieved thanks to gradient timing optimization, ramp sampling and asymmetric echo readouts techniques [54]. Moreover, to allow for partial recovery of the magnetization and to increase the inflow signal of unsaturated blood in the slice being imaged, the 2D image acquisition blocks are interleaved by a waiting time interval. Further details are described in the works of Bauman et al. and Bieri [34,35,54], and typical imaging parameters for ufSSFP are given in the "Pipeline evaluation" section.

MP-MRI feasibility was demonstrated at 0.55T, 1.5T, and 3T, but due to MR hardware specifications at various field strengths and the lung parenchyma physical properties, the best image quality and the highest signal-to-noise ratio (SNR) are currently reached at 1.5T [34,41]. The dedicated MR pulse sequences for MP-MRI provide a higher image quality, SNR, contrast-to-noise ratio (CNR), and reliability of functional maps as compared to standard pulse sequences such as spoiled gradient echo (SPGR) or bSSFP. Specifically, at 1.5T MR field strength, as compared to SPGR sequences which might also be employed for FD functional imaging, the ufSSFP sequence used for MP-MRI delivers an SNR more than three times higher in the lung parenchyma, a CNR 2.5 times higher for ventilation mapping, and a CNR of more than nine times higher for perfusion [35]. Moreover, the ufSSFP, as compared to bSSFP, reduces banding artefacts, increases the pulmonary signal and the quality of the derived functional maps [34]. At 3T, the clear benefit of tSPGR against SPGR and ufSSFP in terms of SNR and CNR was demonstrated by Bauman et al. [35].

### A.2. Data transfer

The datasets are transferred from the PACS or the MR scanner to a local workstation on which TrueLung is installed. The computed functional MR data are tagged with the label "Matrix Pencil

Decomposition", and can be sent back to PACS together with the reports. All the data and summary reports are accessible on the local computer for research purposes.

Optionally, instead of using a local network, the data can be pseudo-anonymized, encrypted, password-protected, and sent via a secure sockets layer (SSL) connection to an external server on which the TrueLung pipeline runs; in this case, at the end of the TrueLung process, the data are sent back securely and can be de-anonymized.

### A.3. Image registration

Precise registration of lung images is a prerequisite for proton-based functional techniques such as MP-MRI. Due to the sliding of the lung in the thoracic cavity and the intensity modulations caused by respiration and blood perfusion, the registration of pulmonary images is challenging. The global smoothness assumptions of the transformations used in common registration algorithms often do not hold for pulmonary imaging due to the local discontinuities between sliding organs, such as the lung and ribcage boundaries. To tackle these issues, an adaptive anisotropic graph diffusion regularization method (GDR) to enforce global smoothness while preserving local discontinuities of the transformation has been developed [42].

Without the need for a lung mask, the GDR registration algorithm preserves discontinuities of the transformation field at sliding organ boundaries, such as between lung and ribcage or liver. At the same time, the algorithm guarantees smoothness in areas with similar motions, such as in the lung. The algorithm preserves cyclic intensity changes caused by the respiratory cycle and by the inflow of blood passing through arteries and parenchymal capillaries at every cardiac pulsation. To reduce computational time, the GDR algorithm is programmed in C++ and CUDA to run on a GPU.

### A.4. Recurrent neural network segmentation

A recurrent neural network (RNN) with main layers consisting of multi-dimensional gated recurrent units (MD-GRUs) is used for voxel-wise binary or multi-class classification [55,56], i.e., whole-lung or lung-lobe segmentation. The network consists of 3 layers of MD-GRUs with 16, 32, and 64 channels, which are connected with pixel-wise fully connected hidden layers of 25 and 45 channels followed by a hyperbolic tangent activation function (tanh). The last MD-GRU is connected to a pixel-wise fully connected layer with $c$ channels, the same number as classes in the data. Finally, the probabilities for each class are estimated using a softmax in the last layer, and consequently, the network is trained by minimizing the multinomial logistic loss, i.e., with a negative log-likelihood cost function. As preprocessing for all the images, a high-pass filter is applied: the Gaussian smoothed images ($\sigma = 5$ pixels) are subtracted from the original images. Both the original and the high-pass filtered images are normalized to $\sigma = 1$ (standard deviation) and $\mu = 0$ (mean), assuming normally distributed values.

During training, an on-the-fly data augmentation is applied to increase the prediction robustness and accuracy, i.e., images and segmentation masks are randomly and slightly scaled, rotated, skewed, distorted, shifted, and noise is added. More details can be found in Andermatt et al. [55,56], and the network implementation is provided at https://github.com/zubata88/mdgru.

The MD-GRU has already shown good accuracy in thoracic and brain segmentation tasks. For whole-lung segmentation, the RNN was initially trained with data of 51 patients examined in previous studies, and incremental learning with additional data from 50 patients was performed in a second moment to improve the robustness of the network. Before the incremental learning, the MD-GRU for whole-lung segmentation reached a dice similarity coefficient of 93.0%, performing comparably well as human observers [44]. Qualitatively the whole-lung segmentations predicted by the RNN appeared well-performed in 94% of slices, and the residual 6% would have required minor manual refinements due to small imperfections such as invasion of lung boundaries (e.g., ribs, chest, bowel), or the disease (e.g., atelectasis, mucus) was partially not included in the lung masks.

The RNN for lobar lung segmentation was trained with MR data from 100 patients employing a cross-modality approach, as explained by Pusterla et al. [45]. For lung-lobe segmentation of ufSSFP data, the network reached an all-lobe Dice similarity coefficient of 93.0±1.8 % (mean±pooled SD) and a median Hausdorff distance of 7±1 mm (mean±SD), indicating good accuracy. The lung-lobe masks were well predicted in 91% of cases, whereas the residual 9% lobe-mask predictions would have required minor refinements due to the mixing of lobe boundaries or lung boundary invasion.

### A.5. The MP algorithm

The MP analysis is performed voxel-wise on registered image series. The signal intensity in the segmented lung is first integrated, and a spectral analysis using matrix pencil is performed in order to detect global signal modulations. The modulations are classified as respiratory or cardiac using k-mean clustering and reference frequency ranges of physiological cycles. The corresponding respiratory and cardiac amplitudes and phases are then retrieved using MP decomposition and least-square linear fitting of the signal modulations, as described in Bauman and Bieri [26]. Lung segmentations of inspiration/expiration images are used to verify fractional ventilation values calculated with MP. The voxels in the lung perfusion maps representing the 95th percentile of signal intensities are identified as reference blood vessels and used to quantify perfusion values, as described by Kjørstad et al. [57]. Further, as a quality check, the perfusion frequencies in the blood vessels are compared to the perfusion frequencies in the whole-lung selected by the MP algorithm to verify proper discrimination. MP offers a fully automated spectral analysis of the time-resolved data sets. Compared to FD, the MP decomposition method improves the signal analysis and mitigates the problems caused by time-series truncation and irregular cardiac frequencies and breathing. This results in an increased accuracy and reproducibility of quantitative parameters derived from the time-resolved ufSSFP acquisitions such as regional fractional ventilation [%], perfusion [ml/min/100ml], and blood arrival time [ms]. The software is implemented in C++ using Armadillo and ITK libraries [58,59].

### A.6. Quantitative evaluation and outcomes

Quantitative evaluation is executed by analyzing the pulmonary signal of functional maps generated using the MP algorithm. To this end, whole-lung or lung-lobe segmentation masks are required. RNN segmentations of the lung parenchyma are performed on the baseline images, excluding large pulmonary vessels. These segmentation masks are applied for pulmonary perfusion quantification. For ventilation quantification, the segmentation masks are refined by removing the vessels as follows: the voxels in the lung perfusion maps representing the 85th percentile of signal intensities are identified as vessels and removed from the segmentation masks for ventilation quantification. Vessels are removed from the ventilation maps since they appear as non-ventilated and must be excluded for ventilation defect quantification [44].

Histogram distribution analysis of perfusion and ventilation values is performed to recover functional defects or hypofunctional lung parts. Depending on the lung pathology investigated and the clinical question, the histogram distribution analysis of functional values can be performed with several algorithms, for example, by defining a cut-off threshold value or linear binning, by clustering the data into distinct groups [19,44,51], or by using artificial neural networks to analyze the whole image at once [49]. Usually, a threshold equal to 60-80% of the median value from each voxel distribution is used to quantify pulmonary regions with impaired relative fractional ventilation ($R_{FV}$) and perfusion ($R_Q$) in every slice and generate ventilation (V) and perfusion (Q) defects maps, as described in previous studies [19,44]. Moreover, maps of ventilation/perfusion defects overlap can be calculated.

The percentage of whole-lung $R_{FV}$, and perfusion $R_Q$ are calculated in every subject as a lung-area-weighted average among the several slices acquired. Similarly, lobar impaired functions are calculated as a lobe-area-weighted average among the acquired slices.

### A.7. Reporting

The TrueLung pipeline outputs in DICOM (Digital Imaging and Communications in Medicine) format and for every slice the registered time series, the baseline image, maps of perfusion, blood arrival time, and fractional ventilation. Moreover, the TrueLung saves maps of pulmonary perfusion and fractional ventilation overlayed on the baseline images, as well as maps of perfusion defects, ventilation defects, and segmentation masks overlayed on the baseline images in the portable network graphics (PNG) format for quick overviews. The whole-lung signal time course analyses are also available in PNG format, along with the whole-lung segmentation of baseline images, the segmentation of vessels, the segmentations of images in inspiration and expiration, and the lung-lobe segmentations (see Figure 2). All the data are easily accessible via a DICOM or PNG viewer and can be further processed and analyzed, for example, in software MITK [60], MATLAB (MathWorks, Natick, MS, US) or Python (Python Software Foundation, Wilmington, Delaware, US).

A PDF summary of the main pulmonary outcomes is created for medical evaluation. The reports include patient and examination data, main outcomes such as percentage of whole-lung defects, and percentage of defects per lung lobe. Maps of ventilation and perfusion, as well as maps of V and Q defects overlaid on baseline images, are also provided. In addition, quantitative value of lung functions such as defects percentage for the whole lung, for every lobe, and for every slice are reported in a tabulated text file. A log-file reports quality-checks warnings, for example caused by unwanted patient movement.

### A.8 Quality control

A review of the TrueLung outcome maps is required since the segmentation masks used for impairment quantification could be imprecisely executed by the RNN. Moreover, it is advised to check the log-file reports and the respective maps in which unwanted patient movement was detected. We account for this quality control with the clinical workflow. Usually, a clinician or radiological technician evaluates both the functional maps and the segmented functional maps; any flaws in the masks or poor functional image quality are easily identified. Segmentation flaws can be quickly fixed manually (e.g. in MITK) [60], and the MP analysis along with the quantitative evaluation module in TrueLung can be rerun (see Figure 2). However, as demonstrated in the result section, manual segmentation refinements are not always necessary for quantification of pulmonary functions. To note, after manual refinement of segmentation masks the MP analysis is also re-executed, but the impact of using precise segmentation masks on the MP analysis, for example on cardiac/breathing frequency selection or quantification of perfusion values, is negligible.

**FIGURES**

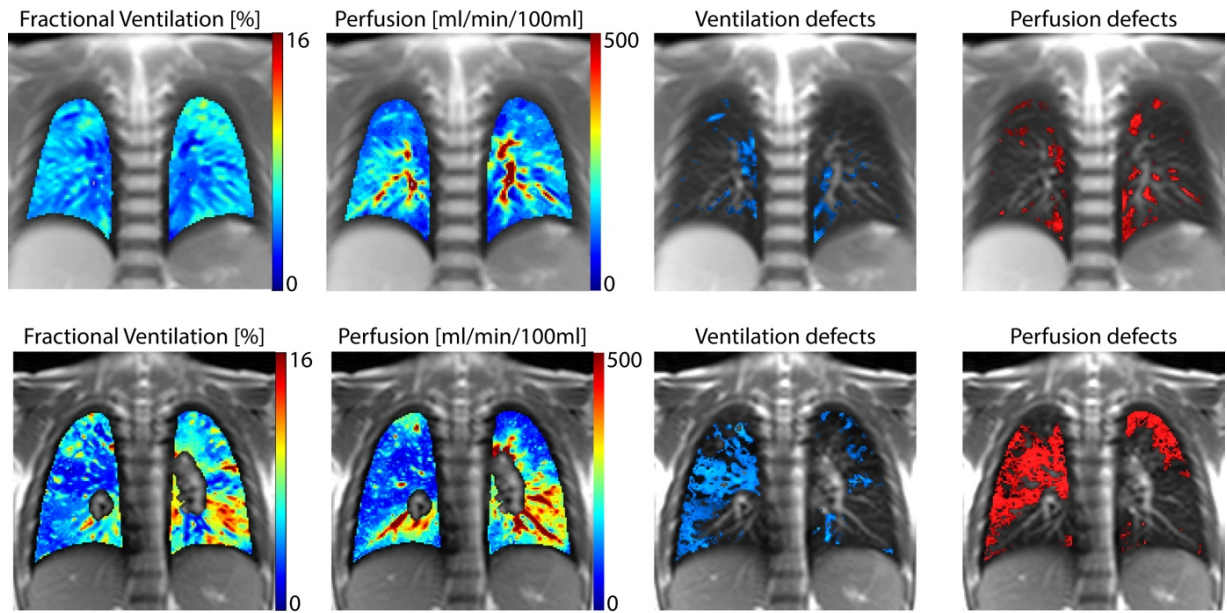

**Figure 1.**

Exemplary functional MRI obtained with MP-MRI in a clinically healthy 11-year-old girl with CF (top row) and in a 16-year-old girl with CF (bottom row) with severe lung disease. In the clinically healthy, fractional ventilation and relative perfusion maps are fairly homogeneous, whereas in the other child, both V and Q maps show large regions with decreased values in both lungs. Masks representing areas with impaired V and Q were overlaid on morphological images: only minor areas with reduced lung function are visible in the first subject, whereas in the other child large functional impairments are visible. To note, the maps of the clinically healthy child show minor functional defects due to the intrinsic inhomogeneity of the lung and the threshold-based quantification method utilized (see Appendix A.6).

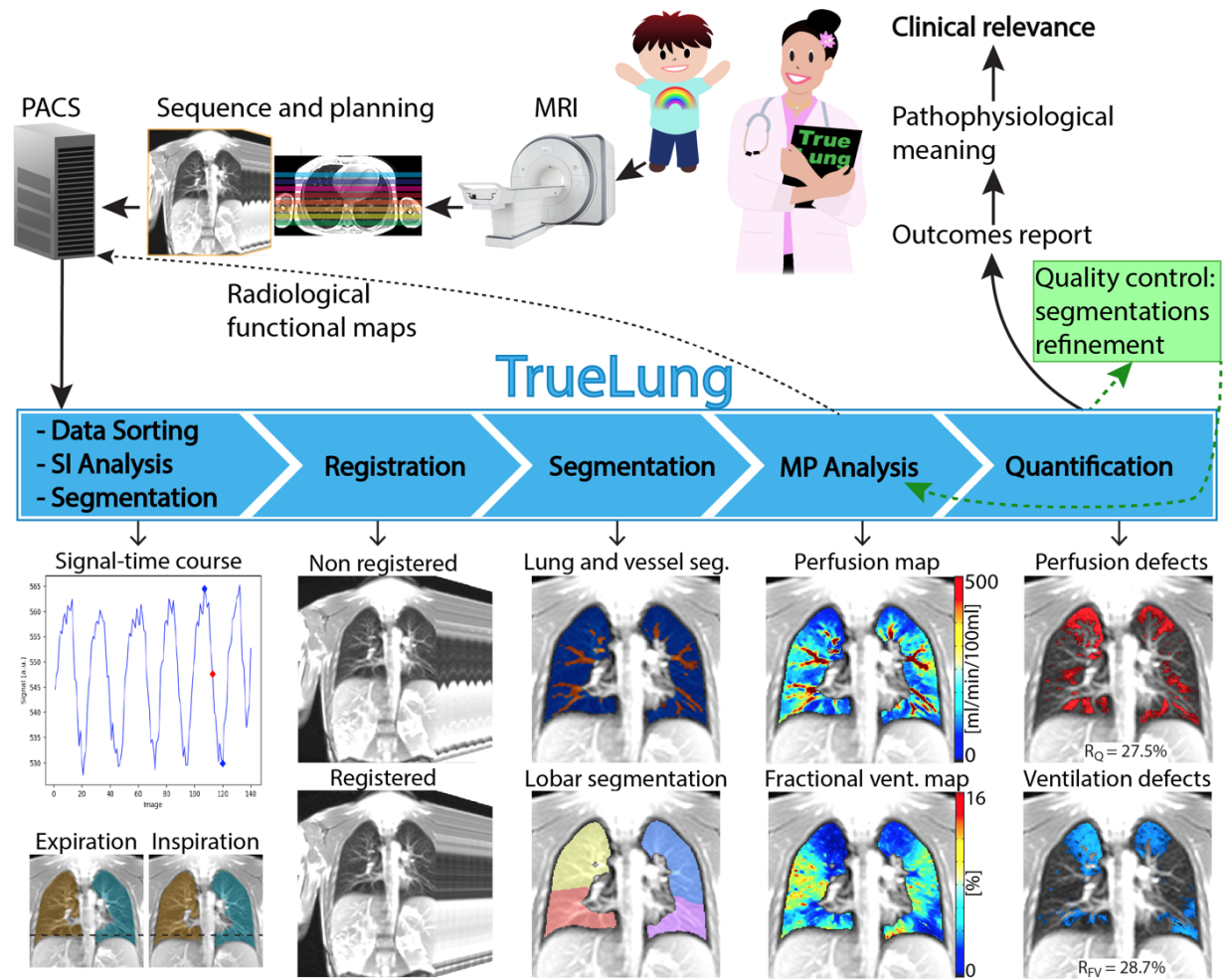

**Figure 2.**

Schematic of the end-to-end workflow, the TrueLung pipeline and its processing steps: data sorting, signal intensity analysis and segmentation of expiratory/inspiratory images, image registration, segmentation of the whole-lung, lung lobes and vessels, MP analysis and generation of functional maps, quantitative defect calculation of $R_{FV}$ and $R_Q$, generation of reports, and quality control. The functional maps after the MP analysis can be sent back to PACS for radiological examination. The quality control includes verification of the segmentation masks and the manual refinement when necessary. Manually refined datasets are re-processed with the MP analysis and the quantification modules. The pipeline can be run on a normal workstation equipped with a CUDA compatible GPU.

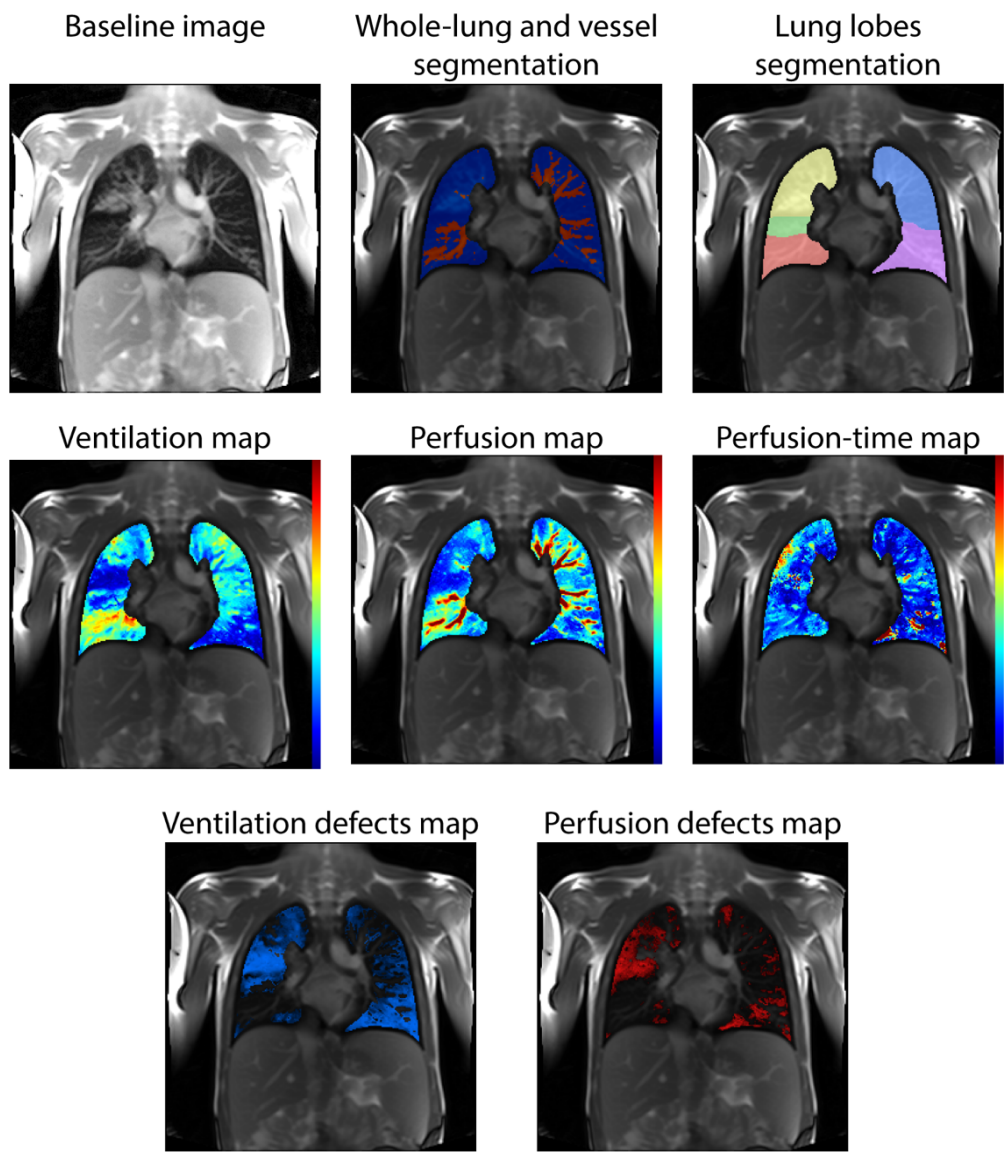

**Figure 3.**

Representative TrueLung output: baseline image, segmentations masks, functional maps and functional defect maps. This 17-year-old CF subject has conspicuous V and Q impairments of the right upper lobe and left lower lobe. The functional maps were segmented and overlaid on the baseline image. Both unsegmented (not shown) and segmented maps can be sent to PACS for radiological interpretation.

**Functional lung MRI report file**

Truelung version: 1.0
Method: MP

Evaluation date: 2022-01-01 15:00
Thresholding: MEDIAN - 75%

**Patient data**

Name: Family Name, Name
Patient ID: 0123456789
Birth date: 2004-01-01
Age: 17 y/o
Sex: M

**Examination data**

Station name: MR1
MR scanner: Aera
Baseline: -

Examination date: 2022-01-01 12:00
Sequence: ufssfp
Study ID: Examination TrueLung

**Global outcomes**

| Function | Slices | Volume [mL] | Defects [%] | Mean ± Std [Units] |
|---|---|---|---|---|
| Ventilation (V) | 11 | 1585 | 31.3 | 6.7 ± 3.9 |
| Perfusion (Q) | 11 | 1585 | 30.9 | 332.3 ± 176.0 |

**Lobar outcomes**

| Function | Lobe | Volume [mL] | Defects [%] | Mean ± Std [Units] |
|---|---|---|---|---|
| Ventilation (V) | LU | 369 | 16.5 | 8.2 ± 3.4 |
| | LL | 372 | 33.7 | 6.3 ± 3.7 |
| | RU | 305 | 49.1 | 5.9 ± 3.8 |
| | RM | 164 | 23.3 | 7.2 ± 3.1 |
| | RL | 374 | 32.4 | 6.4 ± 4.3 |
| | | | | |
| Perfusion (Q) | LU | 369 | 19.6 | 353.8 ± 162.1 |
| | LL | 372 | 20.9 | 366.3 ± 167.6 |
| | RU | 305 | 59.2 | 240.4 ± 147.1 |
| | RM | 164 | 18.9 | 461.0 ± 218.2 |
| | RL | 374 | 34.4 | 310.5 ± 152.5 |

**Ventilation and perfusion maps**

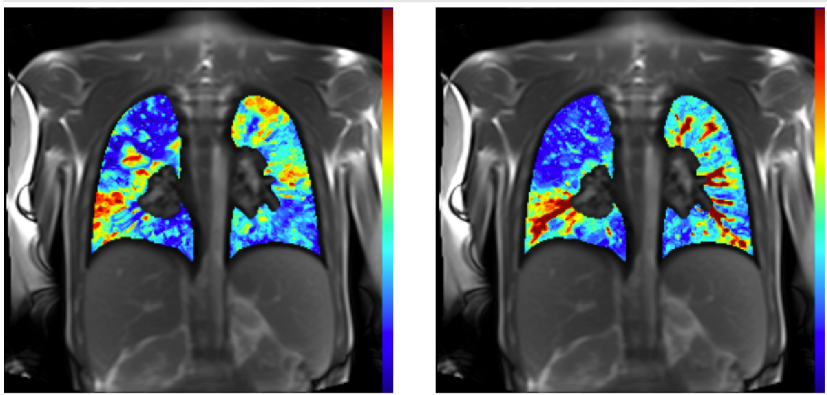

**Ventilation and perfusion impairment maps**

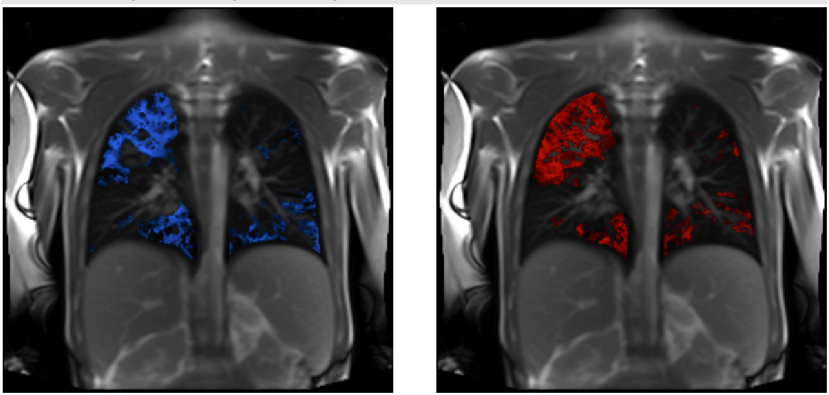

**Figure 4.**

Representative summary report of pulmonary functional MRI provided to clinicians, illustrating the findings for the same subject with cystic fibrosis as presented in Figure 3. For visualization purposes, this report was shortened and presents maps for only one slice. The full report, including all the maps of the slice analyzed, is given in the Supplementary Material S1, available online. The reports include information about the software and execution date, the patient, and examination. Lung functional outcomes, i.e., $R_{FV}$ and $R_Q$ defect percentages, are given in the tables for the whole lung and for every lung lobe. The mean and standard deviation of signal intensity values for ventilation and perfusion are also reported. Ventilation (left) and perfusion maps (right), as well as Q, and V defect masks are included in the report. The CF patient has extended ventilation and perfusion pulmonary impairments. The right upper lobe and left lower lobe are the most affected lung compartments, as visualizable on the maps and demonstrated by the lobar quantitative outcomes. The FEV-1 z-scores in this patient was -2.49 and the LCI was 14.7 turnovers, indicating severe pulmonary disease. As compared to pulmonary function tests, TrueLung allows for regional assessments.

Abbreviations: LU = Left upper lobe; LL = Left lower lobe; RU = Right upper lobe; RM = Right middle lobe; RL = Right lower lobe.

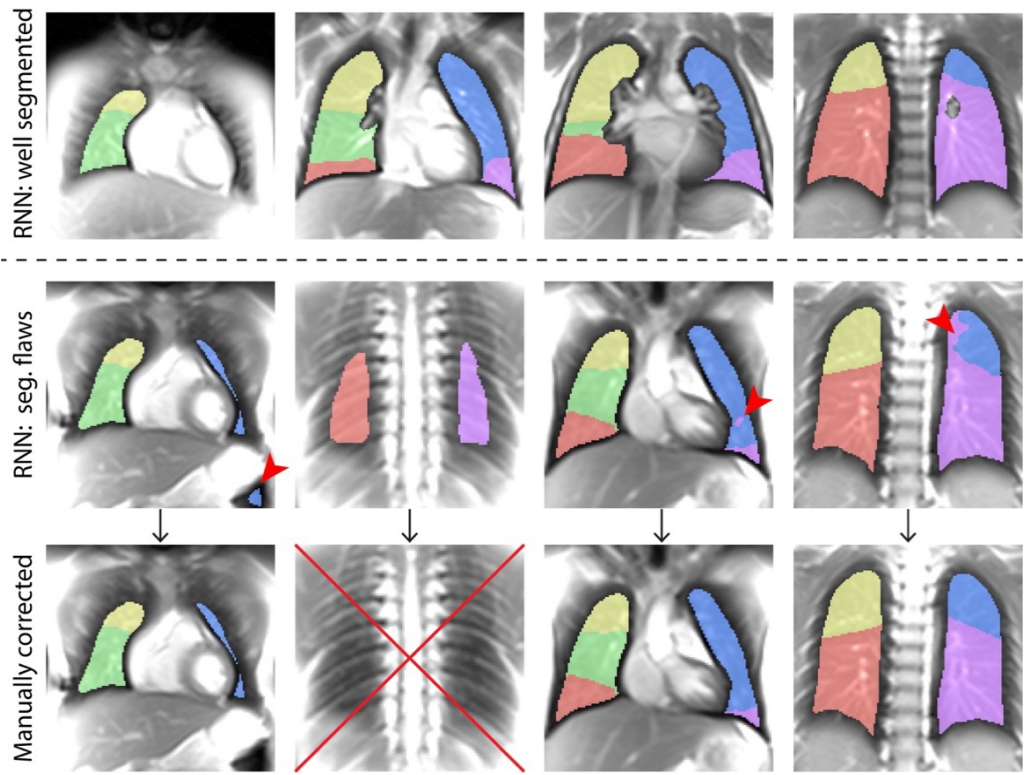

**Figure 5.**

Representative segmentation masks well executed by the RNN (top), imprecisely executed by the RNN (middle row) and successively manually corrected (bottom). The flaws in segmentation masks are indicated by the red arrowheads and were caused by the inclusion of stomach (left column), the inclusion of a very dorsal slice with large partial-volume effects (second column), or "mixing" of lobar boundaries (third and fourth columns). For the sake of clarity, to note the RNN whole-lung segmentation masks (inclusion of all the lobe masks) were wrongly executed for the cases presented in the first and second columns in the middle row, but were well executed for the third and fourth. On the other hand, all the four lobar masks in the middle row required manual corrections and were considered faulty.

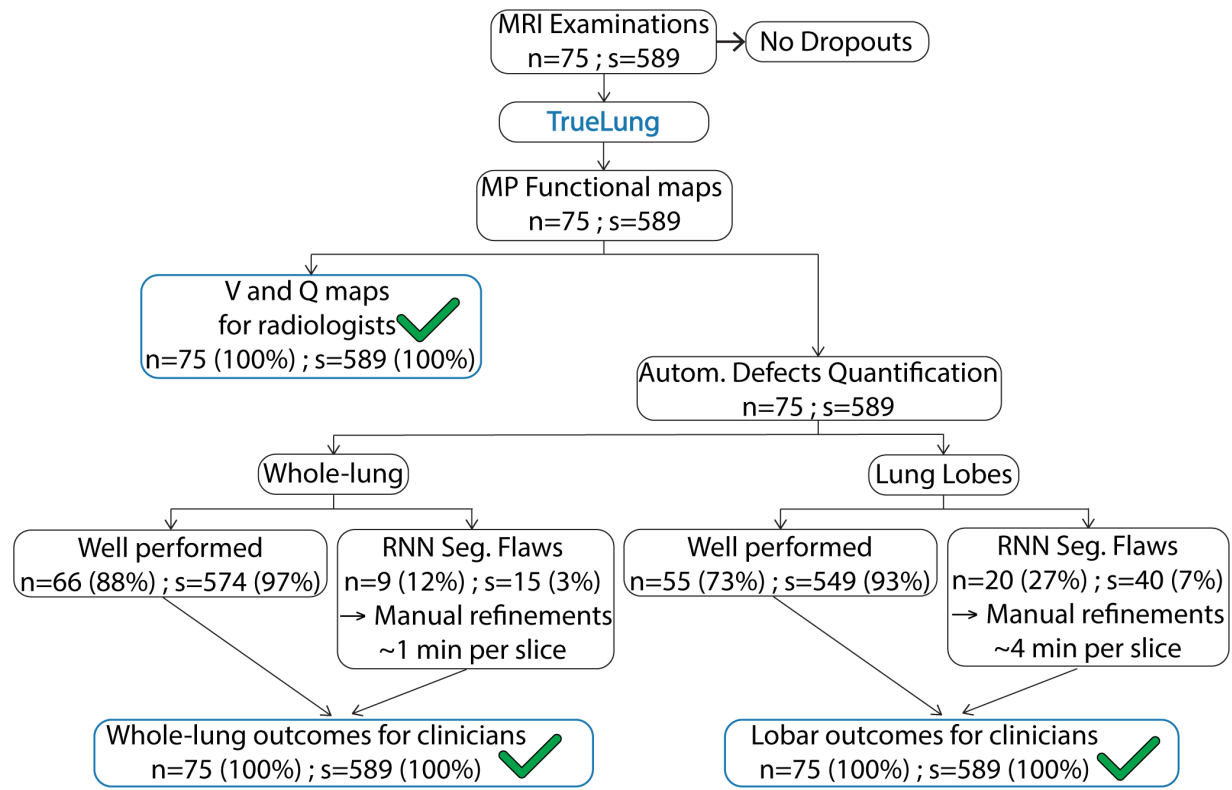

**Figure 6.**

Flowchart of the study. Seventy-five children with CF (n=75) underwent MRI, and all could complete the MR examinations. In total, 589 coronal slices (s=589) were acquired. The 75 examinations were processed with TrueLung. Functional maps were generated for all the 75 examinations and potentially can be delivered to radiologists for diagnosis. The quantitative evaluation of $R_{FV}$ and $R_Q$ defects is performed for the whole-lung and lung lobes, using segmentations masks. For the whole-lung it was well executed for 574 slices, in 66 examinations (left). In nine children, a total of fifteen slices required manual processing due to whole-lung segmentations inaccuracies. The lobar quantification was well executed for 549 slices, in 55 examinations (right). In 20 children, a total of 40 slices required manual refinements. The MP and the quantification module in TrueLung were run again for the manually corrected segmentation masks and whole-lung and lobar $R_{FV}$ and $R_Q$ outcomes of all subjects could be delivered to clinicians.

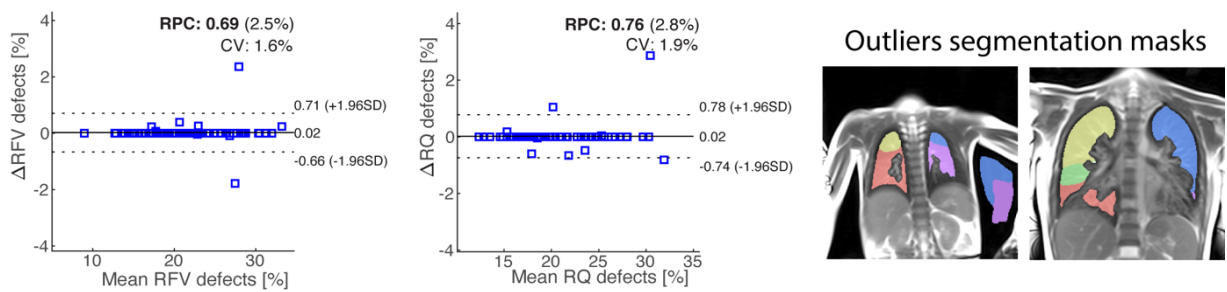

**Figure 7.**

Bland-Altman plots displaying the absolute difference in whole-lung $R_{FV}$ and $R_Q$ defect percentages calculated for fully automated data processed with TrueLung, and data quality controlled and whose segmentations were manually corrected, if needed. The manual correction of lung segmentation masks was performed in 15 slices over 589 (66/75 patients). The impact of the manual segmentation refinement is marginal on the patients $R_{FV}$ and $R_Q$, except for two outliers (see the Bland-Altman fractional ventilation plot on the left), whose segmentation masks are shown on the right. In the Bland-Altman plots the solid line represents the bias, while the dotted lines represent the 95% limits of agreements.

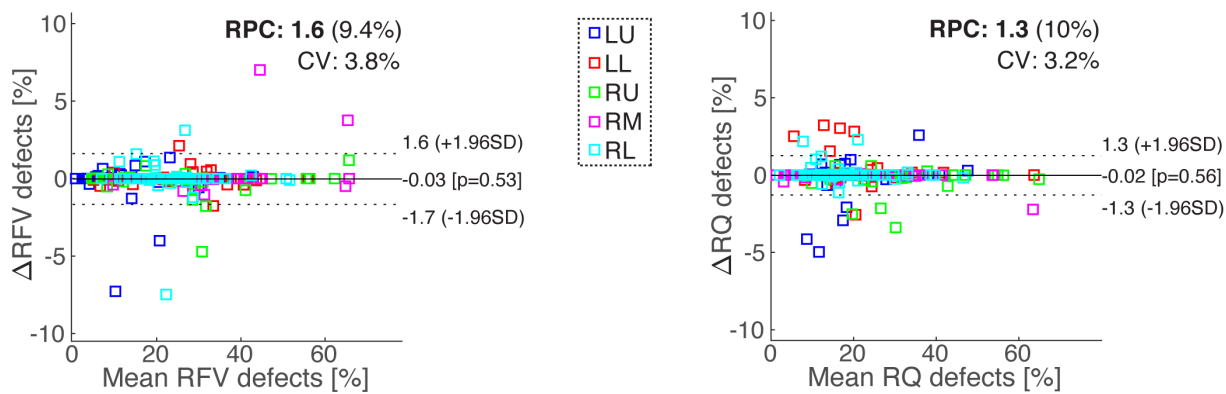

**Figure 8.**

Bland-Altman plots showing the absolute difference in lobar $R_{FV}$ and $R_Q$ defect percentages calculated for fully automated data processed with TrueLung, and data quality controlled and whose lobe segmentations were manually corrected. The manual correction of lobar masks was performed in 40 slices over 589 (55/75 patients). The solid line represents the bias for all the lobes, while the dotted lines represent the 95% limits of agreements. The two outliers previously discussed (cf. Figure 7) were removed from the Bland-Altman analysis. Abbreviations: LU = Left upper lobe; LL = Left lower lobe; RU = Right upper lobe; RM = Right middle lobe; RL = Right lower lobe.

## Supplementary Material S1 and S2

**Supplementary Material S1.** *Entire PDF report for one subject with CF.*

Representative full report of pulmonary functions for the same subject with CF presented in Figures 3 and 4. This report includes maps for all the slices acquired.

# Functional lung MRI report file

Truelung version: 1.0
Method: MP

Evaluation date: 2022-01-01 15:00
Thresholding: MEDIAN - 75%

## Patient data

Name: Family Name, Name
Patient ID: 0123456789
Birth date: 2004-01-01
Age: 17 y/o
Sex: M

## Examination data

Station name: MR1
MR scanner: Aera
Baseline: -

Examination date: 2022-01-01 12:00
Sequence: ufssfp
Study ID: Examination TrueLung

## Global outcomes

| Function | Slices | Volume [mL] | Defects [%] | Mean ± Std [Units] |
|----------|--------|-------------|-------------|--------------------|
| Ventilation (V) | 11 | 1585 | 31.3 | 6.7 ± 3.9 |
| Perfusion (Q) | 11 | 1585 | 30.9 | 332.3 ± 176.0 |

## Lobar outcomes

| Function | Lobe | Volume [mL] | Defects [%] | Mean ± Std [Units] |
|----------|------|-------------|-------------|--------------------|
| Ventilation (V) | LU | 369 | 16.5 | 8.2 ± 3.4 |
| | LL | 372 | 33.7 | 6.3 ± 3.7 |
| | RU | 305 | 49.1 | 5.9 ± 3.8 |
| | RM | 164 | 23.3 | 7.2 ± 3.1 |
| | RL | 374 | 32.4 | 6.4 ± 4.3 |
| | | | | |
| Perfusion (Q) | LU | 369 | 19.6 | 353.8 ± 162.1 |
| | LL | 372 | 20.9 | 366.3 ± 167.6 |
| | RU | 305 | 59.2 | 240.4 ± 147.1 |
| | RM | 164 | 18.9 | 461.0 ± 218.2 |
| | RL | 374 | 34.4 | 310.5 ± 152.5 |

**Ventilation and perfusion maps**

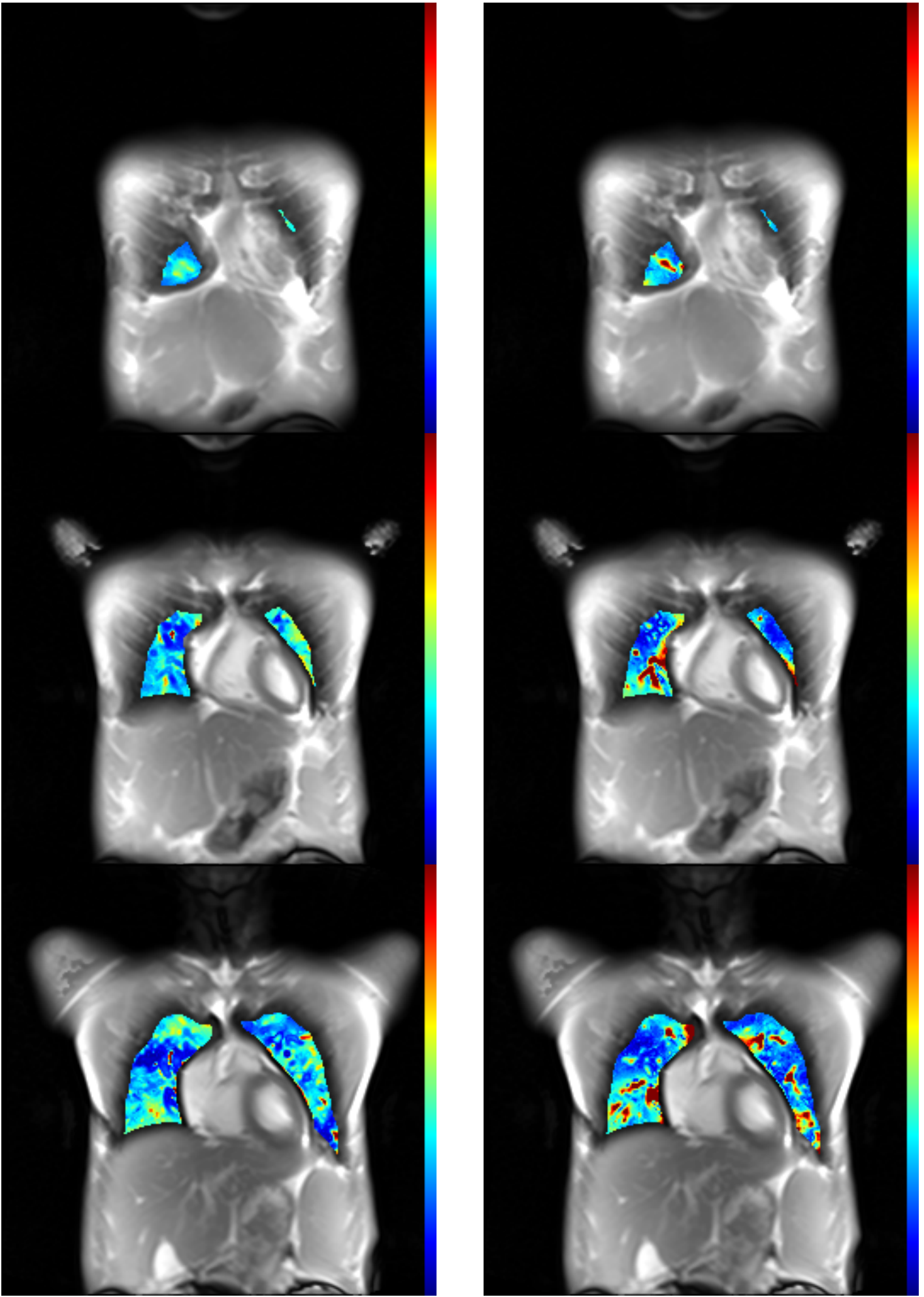

Orso Pusterla et al., *"TrueLung".*

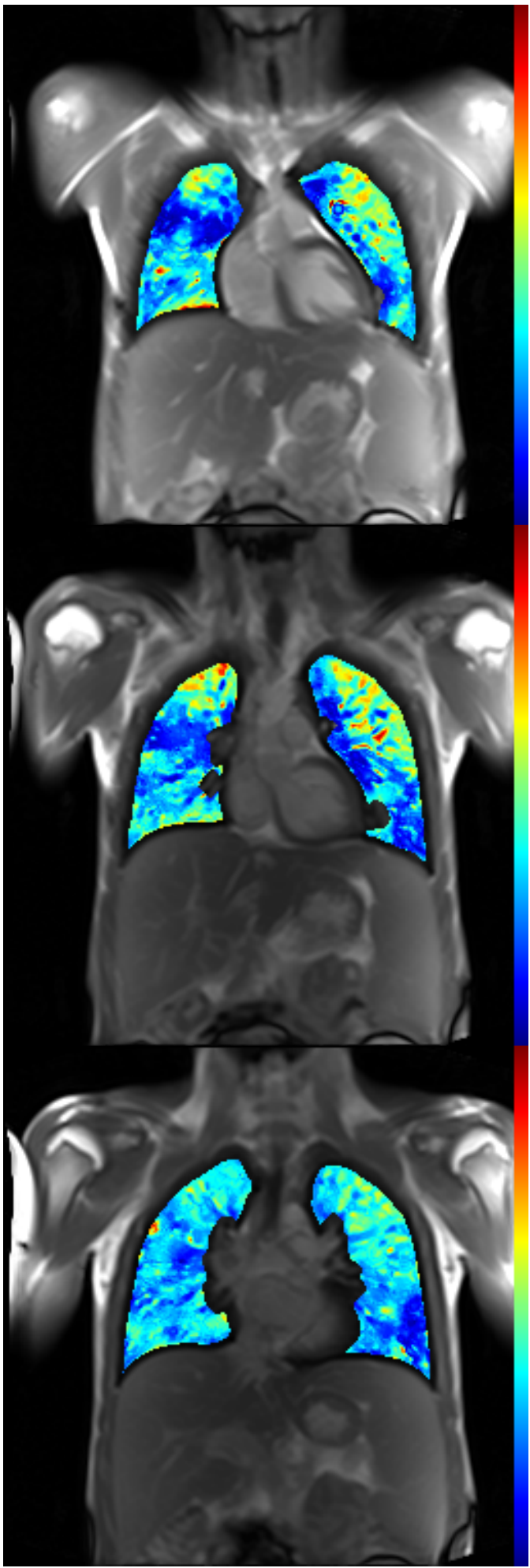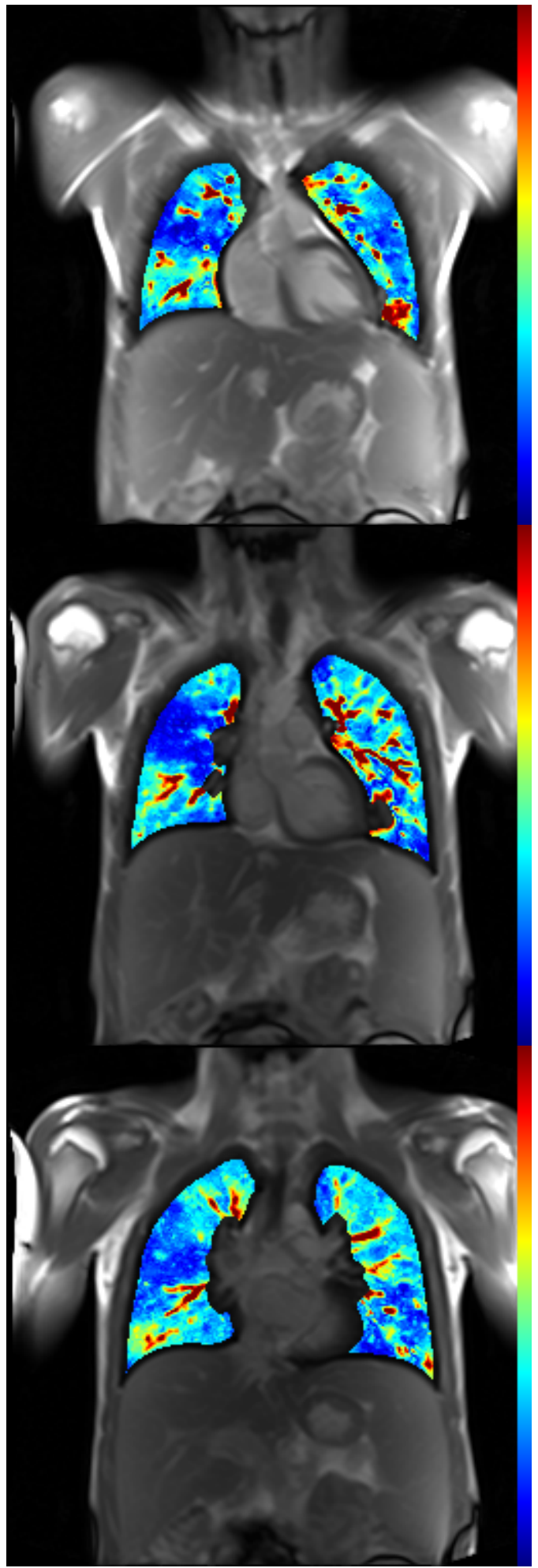

Orso Pusterla et al., *"TrueLung".*

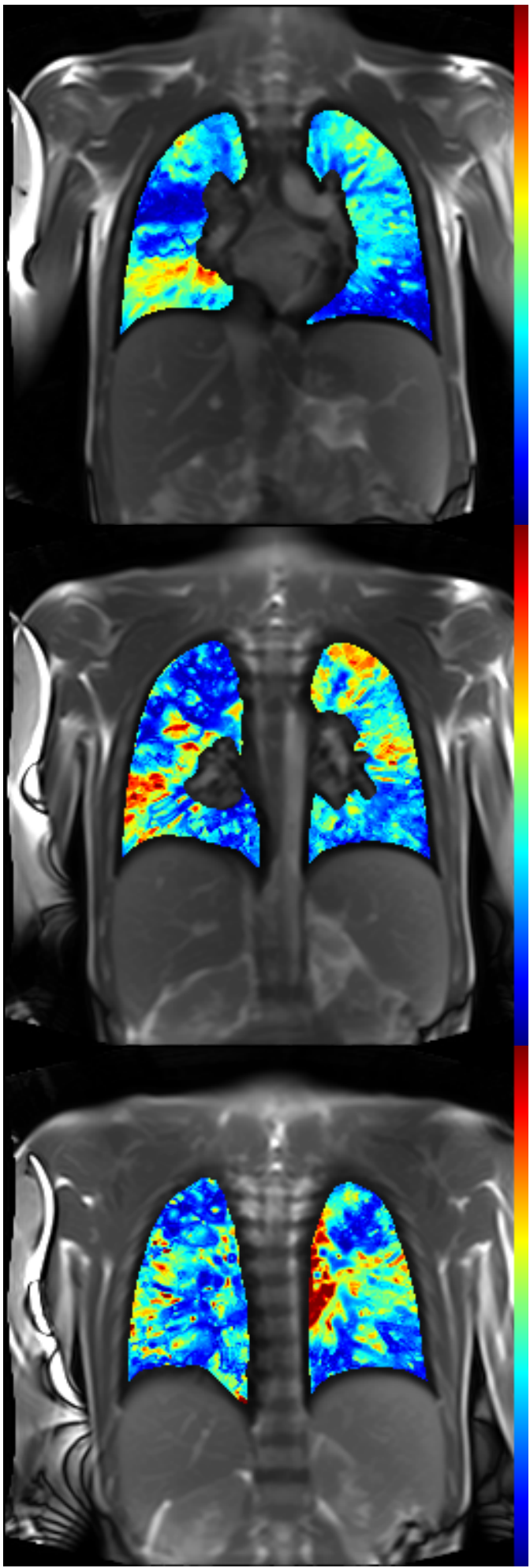
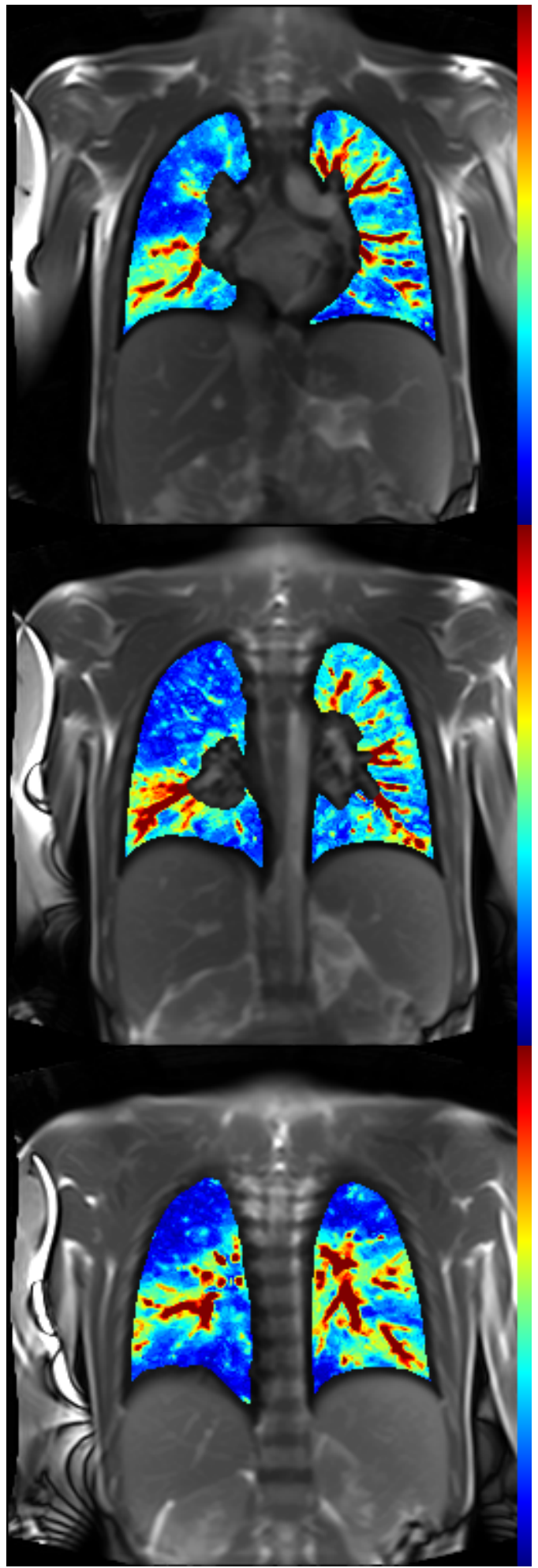

Orso Pusterla et al., *"TrueLung".*

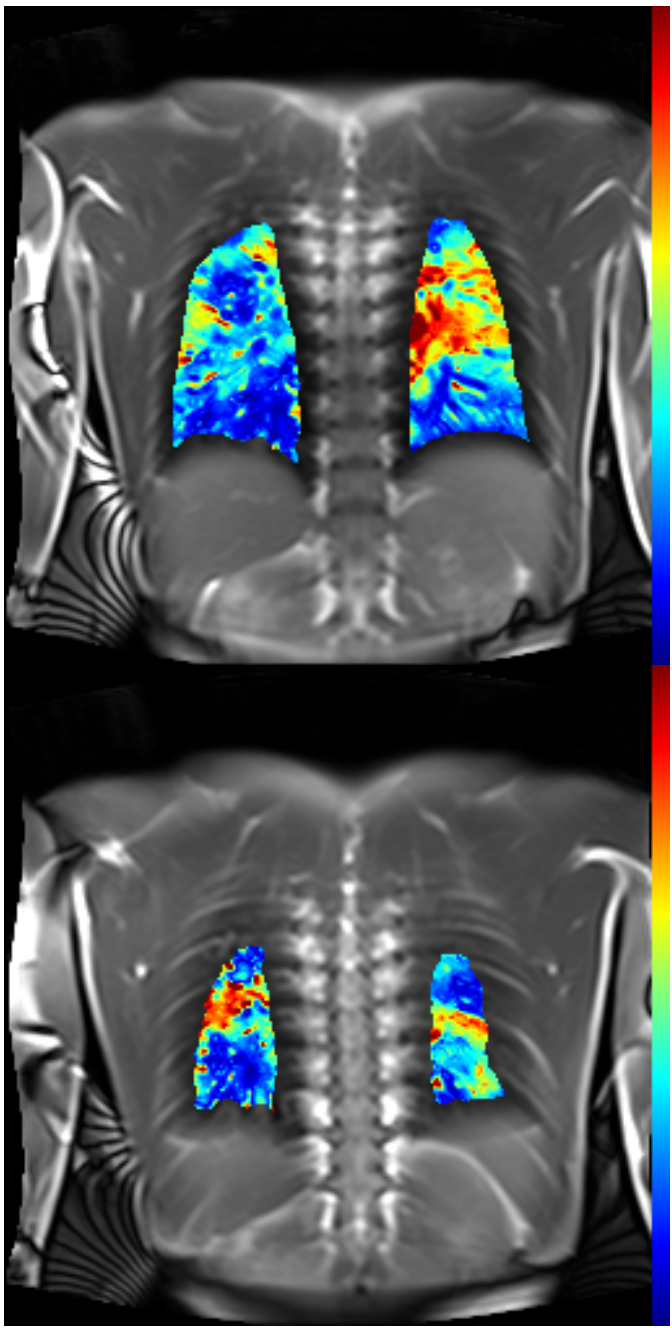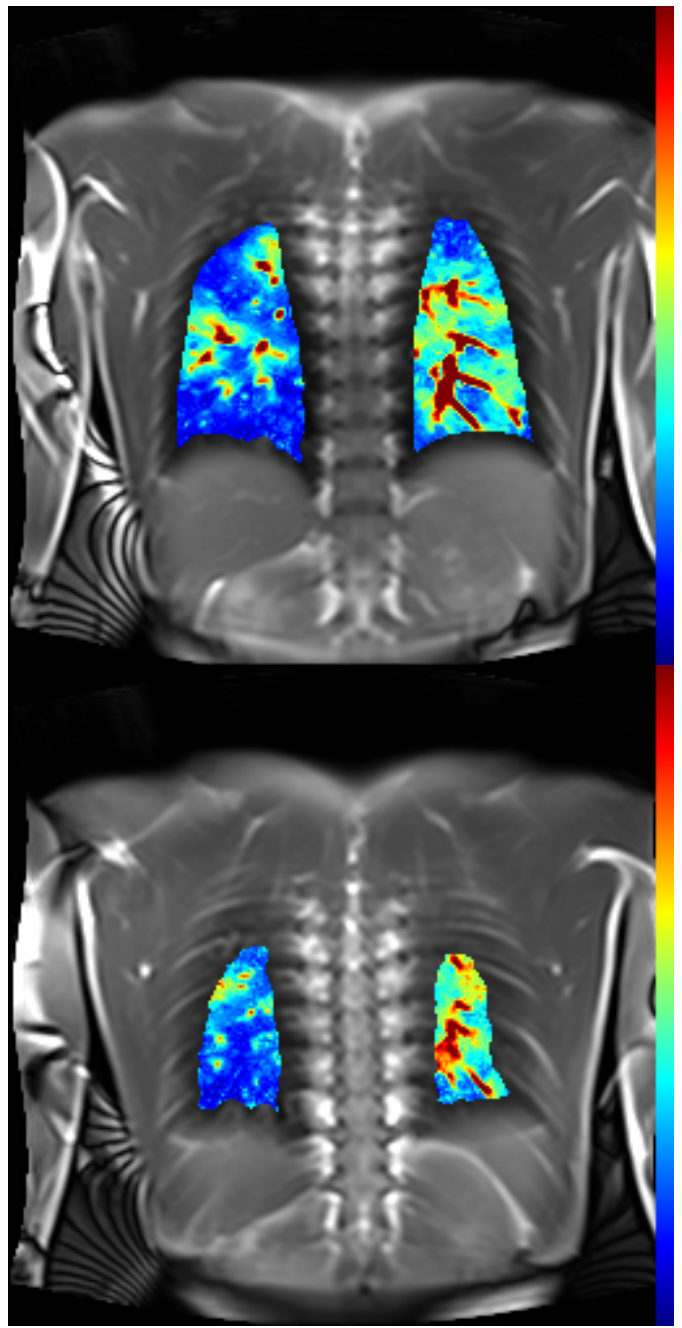

**Ventilation and perfusion impairment maps**

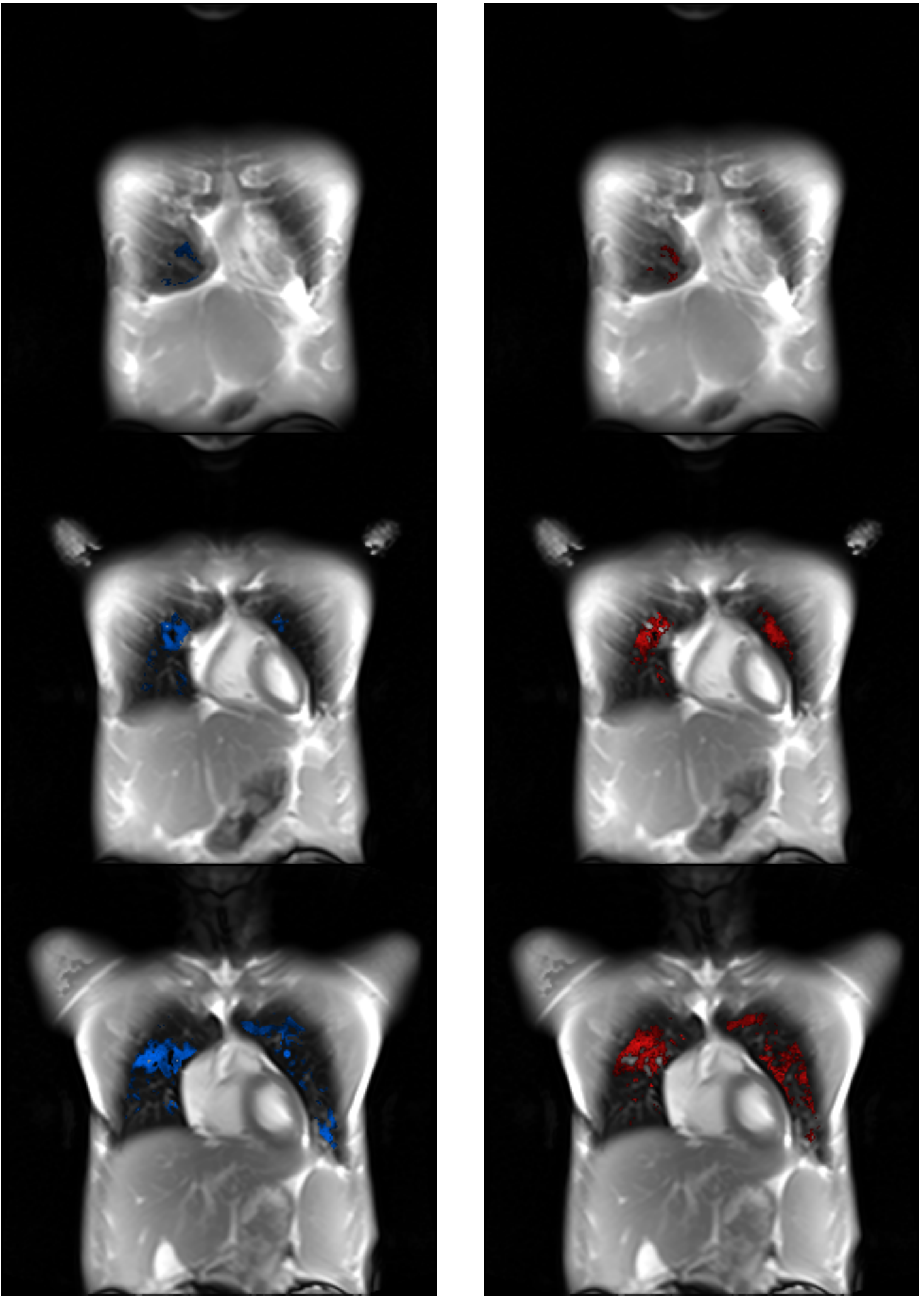

Orso Pusterla et al., *"TrueLung"*.

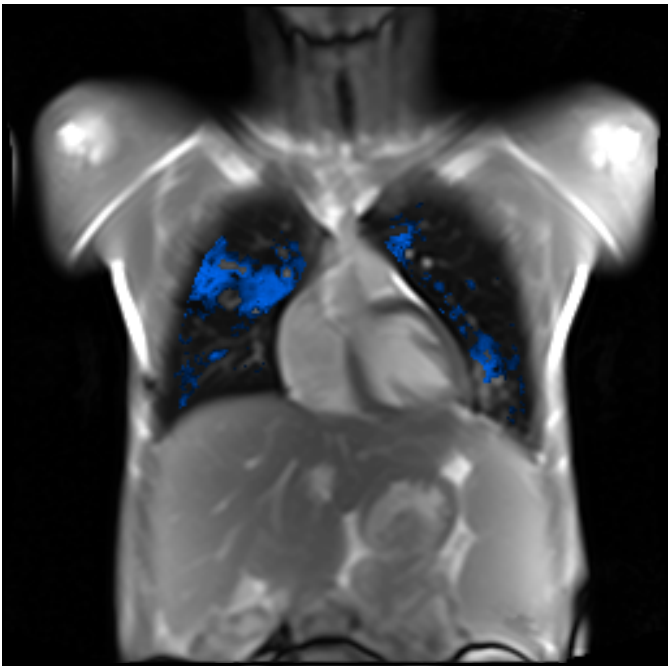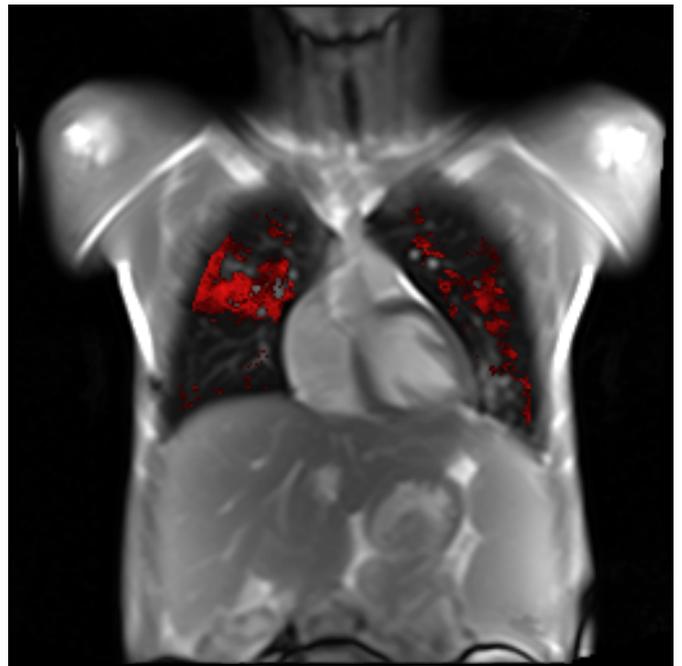

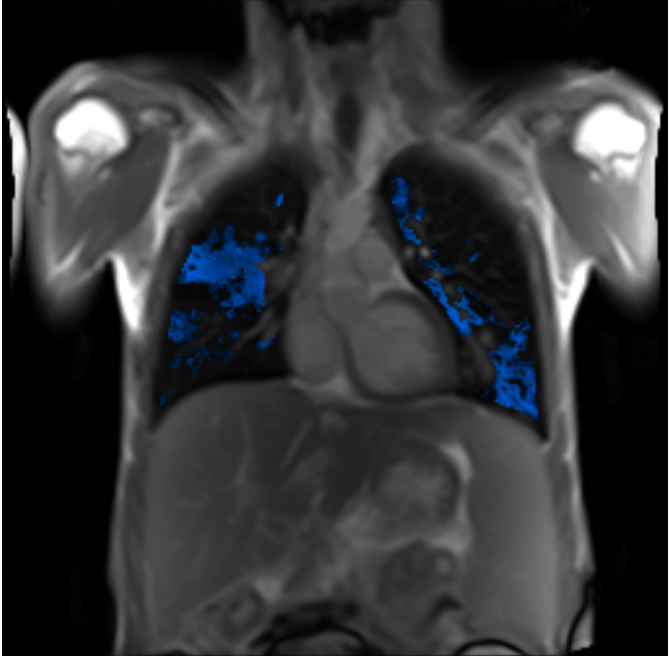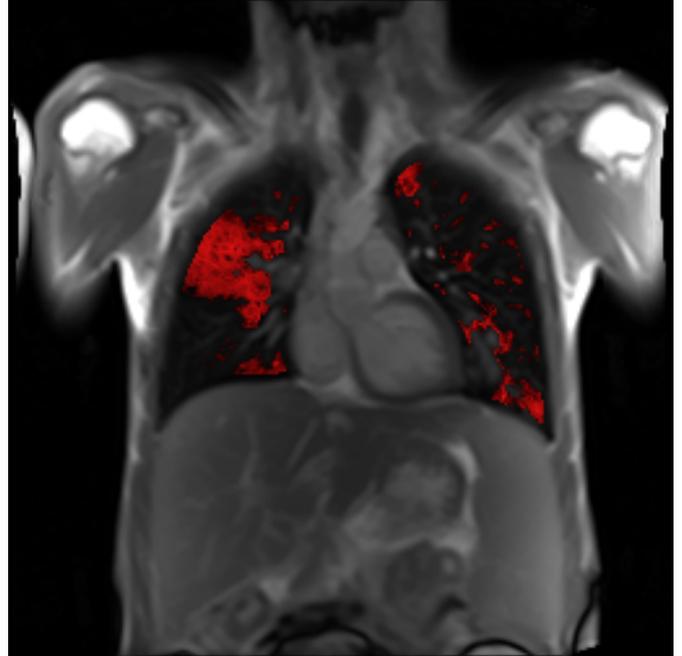

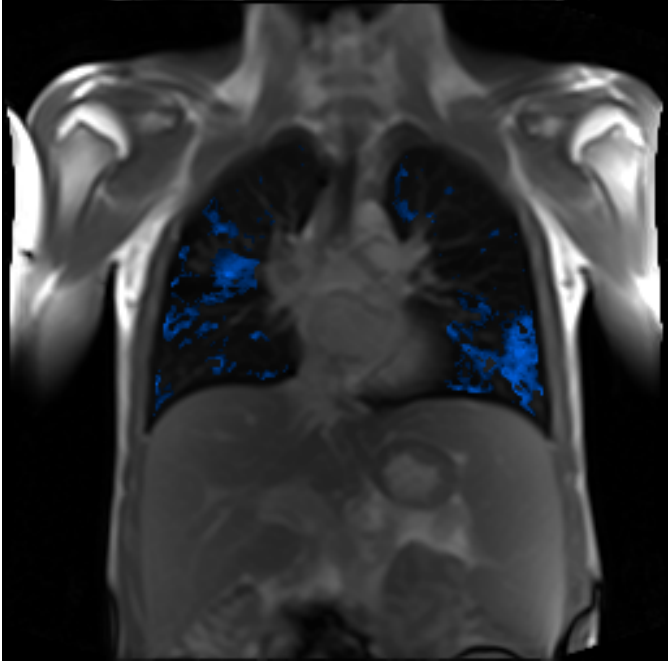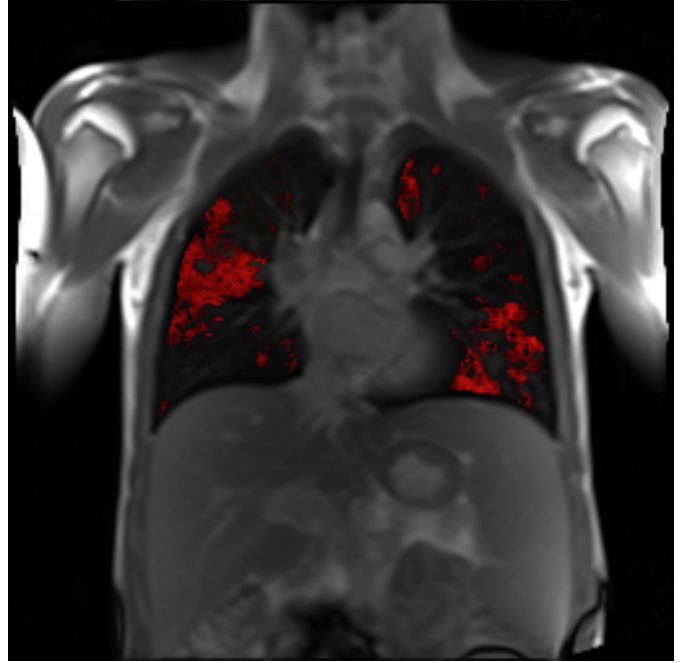

Orso Pusterla et al., *"TrueLung".*

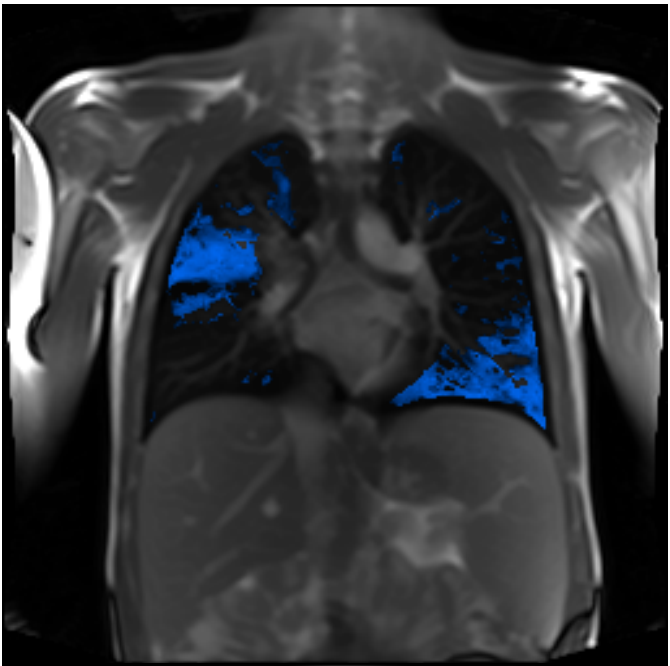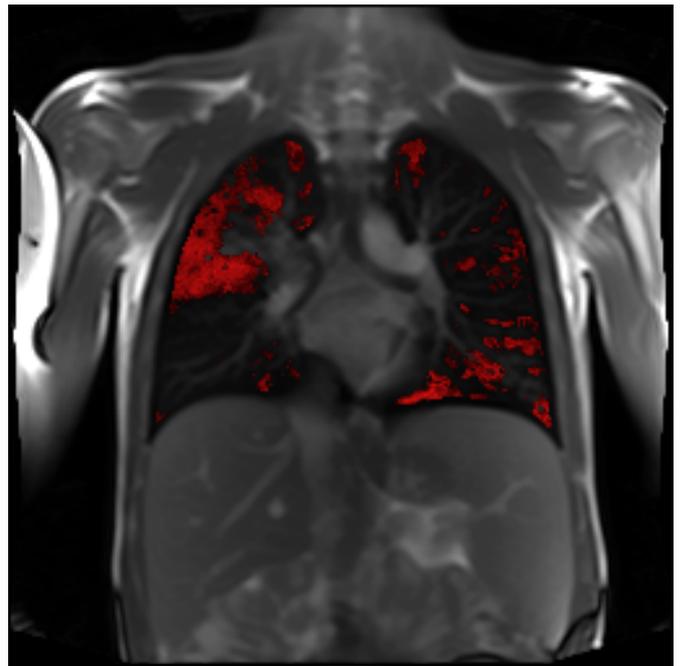

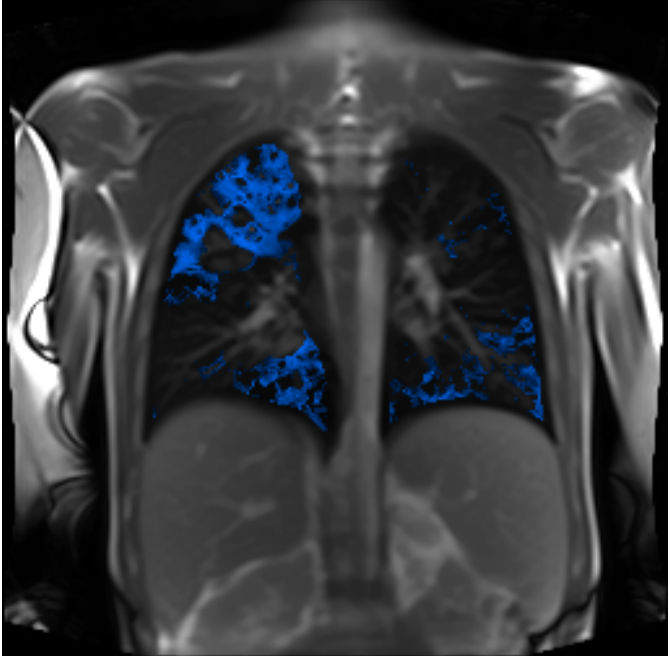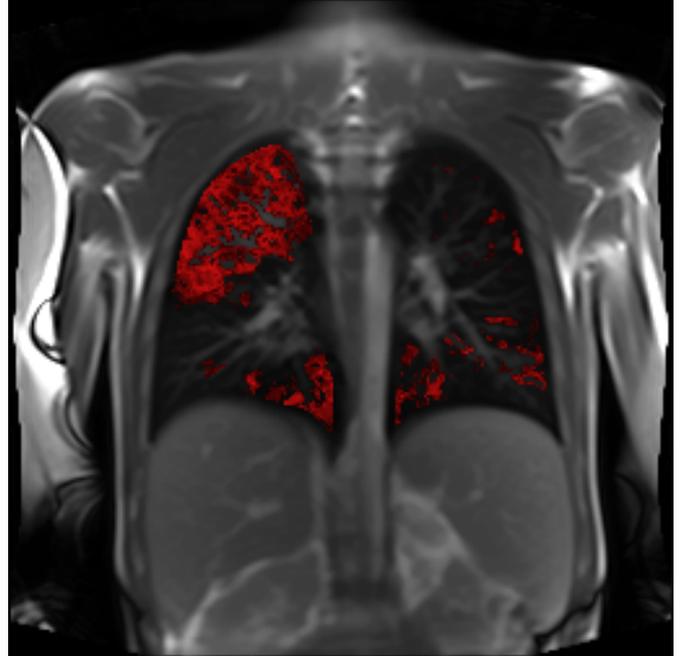

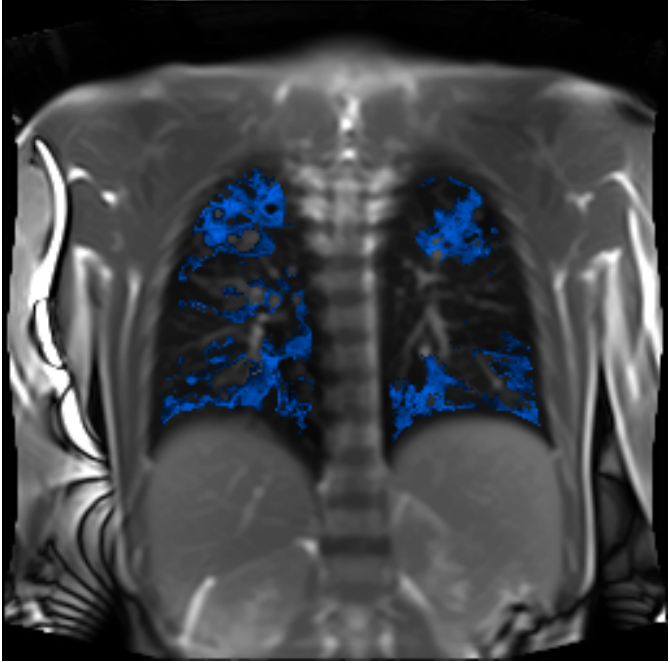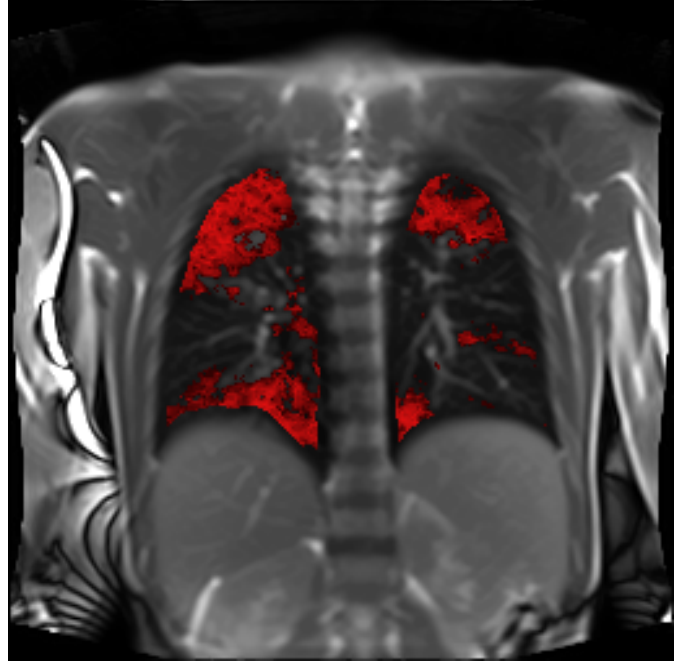

Orso Pusterla et al., *"TrueLung".*

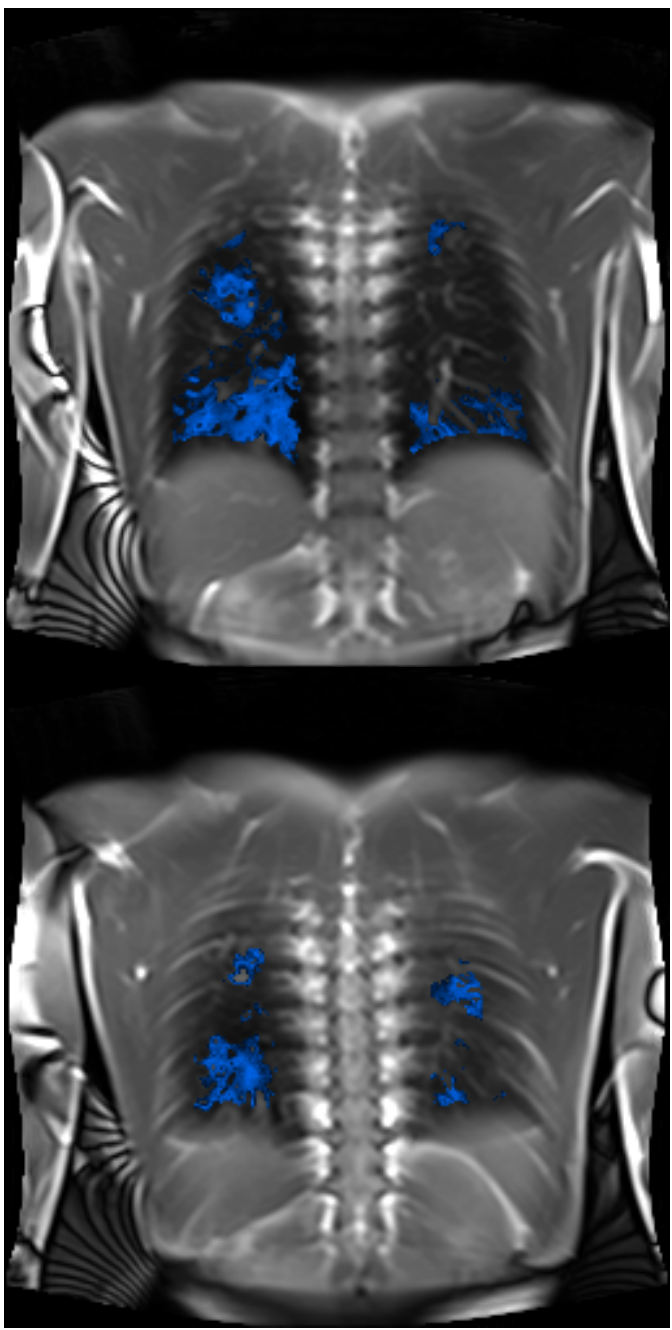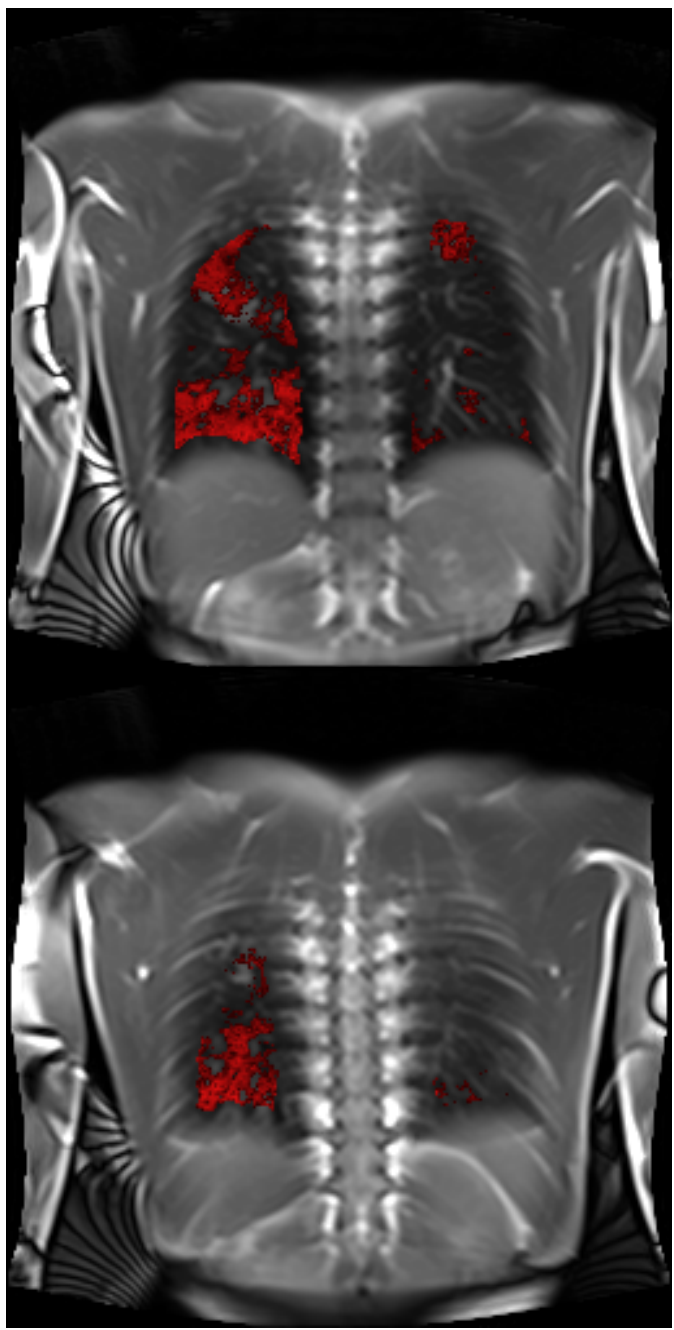

**Supplementary Material S2.** *Extended tabulated pulmonary data file generated with TrueLung.*

The extended tabulated report-file generated by TrueLung gives information for the whole lung, the lobes, and for every acquired slice. The data for the same subject with CF presented in Figures 3 and 4 as well in the Supplementary Material S1 are given.

| Map | N | Volume | Defects | Ratio |
|---|---|---|---|---|
| V_MAP | 11 | 54143 | 16918 | 0.3125 |
| Q_MAP | 11 | 54143 | 16747 | 0.3093 |

| Map | N | Volume | Mean | Stddev | Median |
|---|---|---|---|---|---|
| V_MAP | 11 | 54143 | 6.730 | 3.862 | 6.229 |
| Q_MAP | 11 | 54143 | 332.257 | 176.048 | 297.476 |

| Map | Lobe | Volume | Defects | Ratio |
|---|---|---|---|---|
| V_MAP | LU | 12614 | 2080 | 0.1649 |
| V_MAP | LL | 12719 | 4284 | 0.3368 |
| V_MAP | RU | 10417 | 5110 | 0.4905 |
| V_MAP | RM | 5618 | 1308 | 0.2328 |
| V_MAP | RL | 12775 | 4136 | 0.3238 |
| Q_MAP | LU | 12614 | 2471 | 0.1959 |
| Q_MAP | LL | 12719 | 2659 | 0.2091 |
| Q_MAP | RU | 10417 | 6162 | 0.5915 |
| Q_MAP | RM | 5618 | 1064 | 0.1894 |
| Q_MAP | RL | 12775 | 4392 | 0.3438 |

| Map | Lobe | Volume | Mean | Stddev | Median |
|---|---|---|---|---|---|
| V_MAP | LU | 12614 | 8.190 | 3.410 | 7.896 |
| V_MAP | LL | 12719 | 6.248 | 3.728 | 5.533 |
| V_MAP | RU | 10417 | 5.904 | 3.781 | 5.339 |
| V_MAP | RM | 5618 | 7.206 | 3.056 | 6.890 |
| V_MAP | RL | 12775 | 6.357 | 4.326 | 5.506 |
| Q_MAP | LU | 12614 | 353.831 | 162.049 | 327.530 |
| Q_MAP | LL | 12719 | 366.343 | 167.551 | 352.786 |
| Q_MAP | RU | 10417 | 240.394 | 147.069 | 199.323 |
| Q_MAP | RM | 5618 | 461.016 | 218.185 | 450.757 |
| Q_MAP | RL | 12775 | 310.527 | 152.473 | 276.781 |

| Map | SL | Area | Defects | Ratio |
|---|---|---|---|---|
| V_MAP | 0 | 427 | 102 | 0.2389 |
| V_MAP | 1 | 1469 | 363 | 0.2471 |
| V_MAP | 2 | 3547 | 1051 | 0.2963 |
| V_MAP | 3 | 4624 | 1332 | 0.2881 |
| V_MAP | 4 | 5475 | 1664 | 0.3039 |
| V_MAP | 5 | 6243 | 1552 | 0.2486 |
| V_MAP | 6 | 7168 | 2348 | 0.3276 |
| V_MAP | 7 | 7773 | 2519 | 0.3241 |
| V_MAP | 8 | 8649 | 2913 | 0.3368 |
| V_MAP | 9 | 6107 | 2100 | 0.3439 |
| V_MAP | 10 | 2661 | 974 | 0.3660 |
| Q_MAP | 0 | 427 | 79 | 0.1850 |
| Q_MAP | 1 | 1469 | 486 | 0.3308 |
| Q_MAP | 2 | 3547 | 1093 | 0.3081 |
| Q_MAP | 3 | 4624 | 1302 | 0.2816 |
| Q_MAP | 4 | 5475 | 1621 | 0.2961 |
| Q_MAP | 5 | 6243 | 1655 | 0.2651 |
| Q_MAP | 6 | 7168 | 2128 | 0.2969 |
| Q_MAP | 7 | 7773 | 2508 | 0.3227 |
| Q_MAP | 8 | 8649 | 2982 | 0.3448 |
| Q_MAP | 9 | 6107 | 2044 | 0.3347 |

```
Q_MAP    10    2661      850           0.3194

Map      SL    Area      Mean          Stddev        Median
V_MAP    0     427       8.338         2.003         8.115
V_MAP    1     1469      9.120         3.789         8.815
V_MAP    2     3547      8.403         3.980         8.319
V_MAP    3     4624      6.314         2.880         6.181
V_MAP    4     5475      4.431         1.946         4.268
V_MAP    5     6243      7.176         2.268         7.241
V_MAP    6     7168      9.684         4.924         9.952
V_MAP    7     7773      6.727         3.697         6.064
V_MAP    8     8649      5.954         3.414         5.395
V_MAP    9     6107      6.233         3.831         5.487
V_MAP    10    2661      3.047         2.050         2.606
Q_MAP    0     427       691.824       165.570       673.343
Q_MAP    1     1469      570.111       263.997       534.357
Q_MAP    2     3547      456.621       192.168       433.435
Q_MAP    3     4624      372.252       142.104       361.807
Q_MAP    4     5475      393.611       168.817       396.510
Q_MAP    5     6243      175.353       56.679        176.583
Q_MAP    6     7168      278.756       110.368       265.053
Q_MAP    7     7773      292.124       125.899       277.977
Q_MAP    8     8649      345.485       190.521       315.711
Q_MAP    9     6107      375.966       165.893       360.217
Q_MAP    10    2661      267.888       121.236       246.010

Method    Evaluation    VoxelSize
mp        median 0.75   29.28

Volume        VolumeDiff      Height        Weight
1585.21       202.34          1.61          54

MeanVFreq    StdVFreq    MeanQFreq    StdQFreq
0.360        0.093       1.256        0.082

Truelung Version: 1.0
```